\documentclass[12pt]{amsart}

\usepackage{graphicx,times,amsmath,natbib} 
\usepackage{amsfonts,amssymb}
\usepackage{fullpage,latexsym,epsfig,amsthm}
\usepackage{ifthen,pifont,comment,multirow}
\usepackage{setspace}

\newcommand{\bfy} {\mathbf{y}}

\newcommand{\dd} {\mbox{d}}

\renewcommand{\Pr}{\mathsf{Pr}}
\newcommand{\E}{\mathsf{E}}
\newcommand{\V}{\mathsf{V}}

\newcommand{\reals}{\mathbb{R}}

\DeclareMathOperator{\logit}{logit}
\DeclareMathOperator{\expit}{expit}

\newcommand{\normal}{\mathsf{N}}

\newcommand{\IGam}{\mathsf{IG}}

\newcommand{\bet}{\mathsf{Beta}}

\newtheorem{theorem}{Theorem}

\newtheorem{corollary}{Corollary}
\begin{document}
\title{Stochastic Volatility Models Including Open, Close, High and Low Prices}
\author{Enrique ter Horst \and Abel Rodriguez \and Henryk Gzyl \and German Molina}
\footnote{Enrique ter Horst is Assistant Professor, Instituto de Estudios Superiores de Administraci{\'o}n, Caracas 1010, Venezuela, {\tt enriqueterhorst@gmail.com}.  Abel Rodriguez is Assistant Professor, Department of Applied Mathematics and Statistics, University of California, Santa Cruz, {\tt abel@soe.ucsc.edu}. Henryk Gzyl is Professor, Instituto de Estudios Superiores de Administraci{\'o}n, Caracas 1010, Venezuela, {\tt henryk.gzyl@gmail.com}.  German Molina is Quantitative Trading Researcher at a hedge fund, London, UK, {\tt german@germanmolina.com}.}
\date{}
\begin{abstract}
Mounting empirical evidence suggests that the observed extreme prices within a trading period can provide valuable information about the volatility of the process within that period.  In this paper we define a class of stochastic volatility models that uses opening and closing prices along with the minimum and maximum  prices within a trading period to infer the dynamics underlying the volatility process of asset prices and compares it with similar models that have been previously presented in the literature.  The paper also discusses sequential Monte Carlo algorithms to fit this class of models and illustrates its features using both a simulation study and data form the S\&P500 index.

\end{abstract}

\keywords{Stochastic Volatility; Functionals of Brownian motion; Maximum; Minimum; High; Low; Open; Close; Particle Filter; Bayesian Statistics, Sequential Monte Carlo}

\thanks{}

\maketitle
\section{Introduction}

Understanding the volatility of financial assets is a key problem in econometrics and finance. Over the last two decades, the literature that deals with the modelling of financial asset volatility has expanded significantly. Volatility, a latent variable as such, has been shown to be highly persistent, and all kinds of models have been proposed to extract that persistence and incorporate it into the investment decision-making process, and is of especial interest in the derivatives world, as one of the key inputs for pricing. From the early Autoregressive Conditional Heteroskedastic (ARCH) model of \cite{En82} and the generalized version (GARCH), offered by \cite{Boll86}, to the stochastic volatility approaches \citep{Sh05}, the literature has expanded exponentially, bringing all kinds of adaptations to particular cases (fractional integration, volatility-in-mean equations, etc.), or expansions into the multivariate space and applications in asset management, derivatives pricing and risk management. Most of them share the common feature of taking closing prices over the frequency considered to do model fitting and forecasting.

More recently some structural variations have been proposed. Some of them allow for more flexible uses of the data available. \citep{EngleRussell1998}, use non-equally spaced observations to extract volatility features. \citep{AndBolDiebLab01,BarnShep02} use data at higher frequencies to provide better estimates of the volatility and its persistence at the lower ones. These original papers, and the expansions and advances that followed them, tend to focus on the use of larger amounts of the information available to achieve the common goal of better fitting and forecasting, but still tend to focus their efforts on closing prices at the highest frequency considered, ignoring the paths between those points, even if this information is available.

When it comes to data availability, there have been major changes in the last two decades. From having only (if that) access to daily or weekly closing prices, to having tick-by-tick/bid-ask data available for many series. The increased availability of information must also induce new classes of models and algorithms that tackle that, and that combine the best possible use of that information with an efficient way to process it, so that they remain practical and useful for practitioners. This manuscript tries to achieve these purposes by adding three commonly available series to the statistical process. We continue using closes over the frequency chosen, but add to them the open, high and low data as well, over that period and frequency considered. This data is available for most series through the common data providers (Datastream, Reuters, Bloomberg), and it provides several advantages. First, the use different sources of information (order statistics together with start/end points) about the movement of the underlying, which help to better understand and make inference on its volatility levels and dynamics. The size of the intra-period swings is very informative, and the information content can potentially be very different from that available from the close-to-close data only. Second, adding that information does not have to be a computational burden. We can still provide a quick model that meets the requirements of speed of fitting, while using a larger set of information. Third, because it opens the door to alternative models that could take advantage of the information embedded in those added data features. For example, modifications in the elicitation of the leverage effect in the likelihood, which could potentially be modeled as a relationship between not only return and volatility, but downside range and volatility. Although we do not explore these improvements in this paper, it is worth mentioning that the flexibility provided by the use of this data is there.

The relevance of Close, High, Low and Open (CHLO) data, that is, the relevance of the path followed by the series rather than the start and end points, is well-known. Financial newspapers include it for the frequency reported (daily, weekly, monthly or yearly), data providers add it to their data series, and practitioners in technical analysis have been using it in the past to model volatility by using estimators like ranges or average true ranges \citet{Wilder1978}. Chartists have also been using CHLO data as a key source of information from a graphical/visual standpoint to identify price patterns, trends or reversals.  By assuming that log-prices follow a log-normal diffusion, \citet{Par80} showed that high/low prices provide a highly efficient estimator of the variance compared to estimators based solely on open/close prices. By incorporating information from CHLO prices, \citet{GaKa80} extended the estimator of Parkinson and gained a significant amount of efficiency compared to only including open/close prices. \citet{BallTour84} derived a maximum likelihood analogue of the estimator of Parkinson. In the context of time series modeling of asset return volatility, \citet{GalHsuTauch99} and \citet{AliBraDieb02} take one very relevant step forward by using the range of observed prices (rather than the maximum and minimum directly) to estimate stochastic volatility models. \citet{BraJon05} extend the work in \citet{AliBraDieb02} by including the range and the closing price in the estimation of a stochastic volatility model, but again ignore the actual levels of the maximum and minimum prices.

Full CHLO prices have been used by a number of authors, including \cite{CRogSatch91}, \cite{CRogSatchYoon94} and \cite{CRog98}.  In particular, \citet{MaAt03} derives a maximum likelihood estimator assuming constant volatility, obtaining better performance than existing previous methods on simulated data. Their method, however, is not integrated into a (G)ARCH or stochastic volatility framework, something done by \citet{Lild02} from a maximum likelihood perspective.  We introduce a Bayesian stochastic volatility model that uses full CHLO prices and develop a particle filter approach for infrerence.

Closing prices, especially at the lower frequencies (daily or lower), have been less and less reliable to ascertain realized volatilities based on them, given the patterns recently seen in volumes. For example, the S$\&$P shows unusually large volumes of trades in the last 15 minutes of the sessions. They have been linked to several factors, like high frequency funds unwinding positions accumulated during the day, or exchange-traded funds and hedgers operating in the last minutes to adjust their positions. All of those flows push prices towards outside the extremes observed during the day. However, the actual behavior of the underlying for volatility (risk) purposes is more extreme. Stop-loss/profit-taking levels, as well as positions sizings and their dynamics, are usually defined based on both technical levels and volatility measurements. If those volatilities are based on extremes and paths as well as close-to-close levels, they can potentially lead to very different levels than those based only on close-to-close levels, this applying to any frequency of investment. Intra-day ranges offer, a cleaner, more liquidity-adjusted picture of volatility. Volatility can be very large and still end up with closing prices not far from where they started, while big swings have happened throughout the day. This is especially the case during periods of low liquidity, where the average holding period of any position diminishes significantly, and people react more violently to moves. Stop-loss orders can exacerbate those intra-day swings, while squaring of daily positions for intra-day funds can produce the opposite effect. This is reflective of higher realized volatility and lower liquidity, but it will not be captured by models that only use close-to-close returns.

Our work combines the use of CHLO data with a Bayesian approach, all in a stochastic volatility framework. However, we use a particle filter algorithm to do the filtered estimation, which is computationally quick and can be applied to any frequency of data; which is of special relevance to practitioners handling large amounts of data at the higher frequencies or for processing large numbers of series (or when a quick decision is needed, even if operating with lower frequency data). Section \ref{se:svmodels} provides a quick introduction to the theory behind stochastic volatility models and an introduction to the notation used throughout the paper. Sections \ref{se:extremesv} and \ref{se:inference} present the analytical framework for the joint density of the CHLO prices, the elicitation of prior densities and a description of the particle filtering algorithm used. Sections  \ref{se:missing} and \ref{se:illusa} show how to deal with problems of missing data, and applies our model to weekly CHLO data of the S$\&$P. We also provide a comparison with the stochastic volatility models using only closing prices data and show how different the results can be once added the information contained in the observed extremes. Section \ref{se:ccl} concludes with a summary and a description of potential applications and extensions.

In summary, our net contribution is threefold. First, we provide a coherent model that links the traditional stochastic volatility model with the CHLO data without the need of assumptions beyond those used in that traditional stochastic volatility model. Second, we provide a quick and simple algorithm that allows for fast estimation of this model, which should be very appealing to practitioners. Third, by having changed only the observation equation, we show that these changes can be embedded in any model that uses any other types of evolution equations.

\section{Stochastic volatility models}\label{se:svmodels}

Because of its simplicity and tractability, geometric Brownian motion is by far the most popular model to explain the evolution of the price of financial assets, and has a history dating back to Louis Bachelier  \citep{CoKaBrCrLeLe00}.  A stochastic process $\{ S_t : t \in \reals^{+} \}$ is said to follow a Geometric Brownian motion (GBM) if it is the solution of the stochastic differential equation
\begin{eqnarray}\label{eq:constvar}
\dd S_t & = & \mu S_t \dd t + S_t \sigma \dd B_t
\end{eqnarray}
where $B_t$ is a standard Wiener process and $\mu$ and $\sigma$ are, respectively, the instantaneous drift and instantaneous volatility of process.  GBM implies that the increments of $y_t = \log S_t$ over intervals of the same length are independent, stationary and identically distributed, i.e., $y_{t+\Delta} - y_t = \log S_{t + \Delta} - \log S_t \sim \normal(\Delta\mu, \Delta\sigma^2)$, or equivalently,
$$
S_{t+\Delta} = S_t \exp\left\{ \Delta\mu + \sigma(B(t + \Delta) - B(t)) \right\}
$$

By construction, GMB models assume that the volatility of returns is constant.  However, empirical evidence going back at least to \cite{Ma63}, \cite{Fa65} and \cite{Of73} demonstrates that the price volatility of financial assets tends to change over time and therefore the simple model in \eqref{eq:constvar} is generally too restrictive.  The GBM model can be generalized by assuming that the both  the price and volatility processes follow general diffusions,
\begin{align*}
\dd S_t &= \mu(S_t,\sigma_t) \dd t + \nu(S_t, \sigma_t) \dd W_t \\
\dd \sigma_t &= \alpha(S_t, \sigma_t) \dd t + \beta(S_t, \sigma_t) \dd D_t
\end{align*}

One particularly simple (and popular) version of this approach assumes that, just as before, the price follows a GBM with time varying drift, and that the log-volatility follows an Ornstein-Uhlenbeck (OU) processes \citep{UhlOrn30},
\begin{align}
\dd Y_t & = \mu \dd t + \sigma_t \dd B_t \label{eq:svct_e1}\\
\dd \log(\sigma_t) & = \kappa( \psi - \log(\sigma_t)) + \tau \dd D_t \label{eq:svct_e2}
\end{align}

Practical implementation of these models typically relies on a discretization of the continuos time model in \eqref{eq:svct_e1} and \eqref{eq:svct_e2}.  For the remainder of the paper, we assume that the drift and volatility are scaled so that $\Delta = 1$ corresponds to one trading period (e.g., day or week) and we focus attention on the state-space model
\begin{align}
y_t & = \mu + y_{t-1} + \epsilon^{1}_t &\epsilon^{1}_t &\sim \normal(0, \sigma_t^2) \label{eq:svdt_eq1} \\
\log(\sigma_{t}) &= \alpha + \phi [\log(\sigma_{t-1}) - \alpha]  + \epsilon^{2}_t & \epsilon^{2}_t &\sim \normal ( 0 , \tau^2 ), \label{eq:svdt_eq2}
\end{align}
where $\alpha = \kappa\psi$ and $\phi = 1-\kappa$.  It is common to assume that $0 \le \phi <1$ and $\log(\sigma_0) \sim \normal( \alpha , \tau^2/(1-\phi^2) )$, so that volatilities are positively correlated and the volatility process is stationary (which ensures that the process for the prices is a martingale) with $\alpha$ determining the median of the long-term volatility, $\nu = \exp\{ \alpha \}$.  Therefore, we expect the volatility of returns to take values greater than $\nu$ half of the time, and viceversa.

Unlike ARCH and GARCH models \citep{En82,Boll86}, where a single stochastic process controls both the evolution of the volatility and the observed returns, stochastic volatility models use two coupled processes to explain the variability of the returns.  By incorporating dependence between $\epsilon^{1}_t$ and $\epsilon^{2}_t$, the model can accommodate leverage effects, while additional flexibility can be obtained by considering more general processes for the volatility (for example, higher order autoregressive process, jump process, and linear or nonlinear regression terms).

Although theoretical work on stochastic volatility models goes back at least to the early 80s, practical application was limited by computational issues.  Bayesian fitting of stochastic volatility models has been discussed by different authors, including \cite{JaPoRo94} and \cite{KiShCh98}.
Popular approaches include Gibbs sampling schemes that directly sample from the full conditional distribution of each parameter in the model, algorithms based on offset mixture representations that allow for joint sampling of the sequence of volatility parameters, and particle filter algorithms.
In the sequel, we concentrate on sequential Monte Carlo algorithms for the implementation of stochastic volatility models.

\section{Incorporating extreme values in stochastic volatility models}\label{se:extremesv}

\subsection{Joint density for the closing price and the observed extremes of geometric Brownian motion}

The goal of this section is to extend the stochastic volatility model described in the previous Section to incorporate information on the full CHLO prices.  To do so, note that the Euler approximation in \eqref{eq:svdt_eq1} implies that, conditionally on the volatility $\sigma_t$, the distribution of the increments of the asset price follows a Geometric Brownian motion with constant volatility $\sigma_t$ during the $t$-th trading period.  Therefore, the discretization allows us to interpret the process generating the asset prices as a a sequence of conditionally independent processes defined over disjoint and adjacent time periods; within each of these periods the price process behaves like a GBM with constant volatility, but the volatility is allowed to change from period to period.  This interpretation of the discretized process is extremely useful, as it allows us to derive the joint distribution for the closing, maximum and minimum price of the asset within a trading period using standard stochastic calculus tools.
\begin{theorem}\label{thm:1}
Let $Y_t$ be a Brownian motion with drift and consider the evolution of the process over a time interval of unit length where  $Y_{t-1}$ and $Y_{t}$ are the values of the process at the beginning and end of the period, and let $M_t = \sup_{t-1 \le s \le t} \{Y_s\}$ and $m_t = \inf_{t-1 \le s \le t} \{Y_s\}$ be, respectively, the supremum and the infimum values of the process over the period.  If we denote by $\mu$ and $\sigma_t$ the drift and volatility of the process between $t-1$ and $t$ (which are assumed to be fixed within this period), the joint distribution of $M_t$, $m_t$ and $Y_t$ conditional on $Y_{t-1} = y_{t-1}$ is given by
\begin{eqnarray*}
\Pr(m_t \ge a_t, M_t \le b_t, Y_t \le c_t | Y_{t-1} = y_{t-1}) & = & \int_{-\infty}^{c_t} q(y_{t},a_t,b_t | y_{t-1}) \dd y_t
\end{eqnarray*}
for $m_t \le \min\{ c_t, y_{t-1} \}$, $M_t \ge \max \{ c_t, y_{t-1} \}$ and $m_t \le M_t$, where
\begin{equation}\label{eq:jointcdf}
\begin{aligned}
q(y_{t},a_t,b_t | y_{t-1}) &=  \frac{1}{\sqrt{2\pi}\sigma_t}  \exp\left\{  -\frac{[ \mu^2-2\mu (y_{t} - y_{t-1}) ]}{2\sigma_t^2}  \right\} \sum_{n=-\infty}^\infty \left( \exp\left\{-d_1(n)\right\}  - \exp\left\{-d_2(n) \right\} \right)
\end{aligned}
\end{equation}
and
\begin{align*}
d_1(n) & = \frac{[y_{t}-y_{t-1}-2n(b_t-a_t)]^2}{2\sigma_t^2} & d_2(n) & = \frac{[y_{t}+y_{t-1}-2a_t-2n(b_t-a_t)]^2}{2\sigma_t^2}
\end{align*}
\end{theorem}

The proof of this theorem, which is a simple extension of results in \cite{Fe71}, \cite{DyMc85} and \cite{Kl05}, can be seen in Appendix \ref{ap:th1}.  An equivalent but more involved expression was obtained by \cite{Lild02}, who used it to construct GARCH models that incorporate information on CHLO prices.  Note that if we take both $a_{t} \to -\infty$ and $b_{t} \to \infty$, the cumulative distribution in \eqref{eq:jointcdf} reduces to the integral of a Gaussian density with mean $y_{t-1} + \mu$ and variance $\sigma_t^2$, which agrees with \eqref{eq:constvar}.

\begin{corollary}
The joint density for $m_t, M_t, Y_t | Y_{t-1} = y_{t-1}$ is given by
\begin{equation}\label{eq:jointpdf}
\begin{aligned}
p(a_t, b_t, y_t | y_{t-1}) &= - \frac{\partial^2 q}{\partial a_{t} \partial b_{t}} q(y_{t},a_t,b_t | y_{t-1}) \\
  &= \frac{1}{\sqrt{2\pi}\sigma_{t}^{3}} \exp\left\{  -\frac{\mu^2-2\mu (y_{t} - y_{t-1})}{2\sigma_t^2}   \right\} \times \\
  & \;\;\; \sum_{n=-\infty}^{\infty} \left[ 4n^2 \left(2d_1(n)-1 \right) \exp\left \{ - d_1(n) \right\} - 4n(n-1) \left( 2d_2(n)-1) \right) \exp\left \{ - d_2(n) \right\} \right]
\end{aligned}
\end{equation}
for $m_t \le \min\{ y_t, y_{t-1} \}$, $M_t \ge \max \{ y_t, y_{t-1} \}$ and $m_t \le M_t$, and zero otherwise.
\end{corollary}

Equation \eqref{eq:jointpdf} provides the likelihood function for the closing, maximum and minimum prices given the volatility and drift, under the first order Euler approximation to the system of stochastic differential equations in \eqref{eq:svct_e1} and \eqref{eq:svct_e2}.  Therefore, the basic underlying assumptions about the behavior of asset prices are the same as in standard volatility models; however, by employing  \eqref{eq:jointpdf} instead of \eqref{eq:svdt_eq1} we are able to coherently incorporate information about the observed price extremes in the inference of the price process.  After a simple transformation $r_t=b_t-a_t$ and $w_t=a_t$ and marginalization over $w_t$, we recover the likelihood function described in \cite{BraJon05},
\begin{equation}\label{eq:rangeclosepdf}
\begin{aligned}
p(r_t, y_t | y_{t-1}) &= \frac{1}{\sqrt{2\pi}\sigma_t} \exp\left\{ -\frac{(y_t - y_{t-1} - \mu)^2}{2\sigma_t^2} \right\} \\
& \;\;\;\; \sum_{n=-\infty}^{\infty} \left[ \frac{4n^2}{\sqrt{2\pi}\sigma_t^2} \left\{ \frac{(2nr_t - |y_{t} - y_{t-1}|)^2}{\sigma_t^2} - 1 \right\} \exp\left\{  \frac{(2nr_t - |y_t - y_{t-1}|)^2}{2\sigma_t^2} \right\} \right.  \\
& \;\;\;\;\;\;\; + \frac{2n(n-1)}{\sqrt{2\pi}\sigma_t^2} (2nr_t - |y_{t} - y_{t-1}|)  \exp\left\{  \frac{(2nr_t - |y_t - y_{t-1}|)^2}{2\sigma_t^2} \right\} \\
& \;\;\;\;\;\;\; \left. + \frac{2n(n-1)}{\sqrt{2\pi}\sigma_t^2} (2(n-1)r_t + |y_{t} - y_{t-1}|)  \exp\left\{  \frac{(2(n-1)r_t + |y_t - y_{t-1}|)^2}{2\sigma_t^2} \right\} \right]
\end{aligned}
\end{equation}
for $r_t > |y_{t} - y_{t-1}|$, while a further marginalization over the closing price $y_t$ yields the (exact) likelihood underlying the range-based model in \cite{AliBraDieb02}, which is independent of the opening price $y_{t-1}$,
\begin{equation}\label{eq:rangeonlypdf}
\begin{aligned}
p(r_t) &= 8 \sum_{n=1}^{\infty} (-1)^{n-1} \frac{n^2 r_t}{\sqrt{2 \pi}\sigma_t} \exp\left\{ -\frac{n^2 r_t^2}{2\sigma_t^2} \right\}
\end{aligned}
\end{equation}

Using either \eqref{eq:rangeclosepdf} or \eqref{eq:rangeonlypdf} as likelihoods entails a loss of information with respect to the full joint likelihood in  \eqref{eq:jointpdf}.  In the case of \eqref{eq:rangeonlypdf}, the effect is clear as the range is an ancillary statistic for the drift of the diffusion, and therefore provides no information about it.  This leads \cite{AliBraDieb02} to assume that the drift is zero, which has little impact in the estimation of the volatility of the process, but might have important consequences for other applications of the model, such as option pricing.  In the case of \eqref{eq:rangeclosepdf}, although there is information about the drift contained in the opening and closing prices, the model ignores the additional information about the drift contained in the actual levels of the extremes.

\begin{figure}
\begin{center}
\includegraphics[height=3.2in,angle=0]{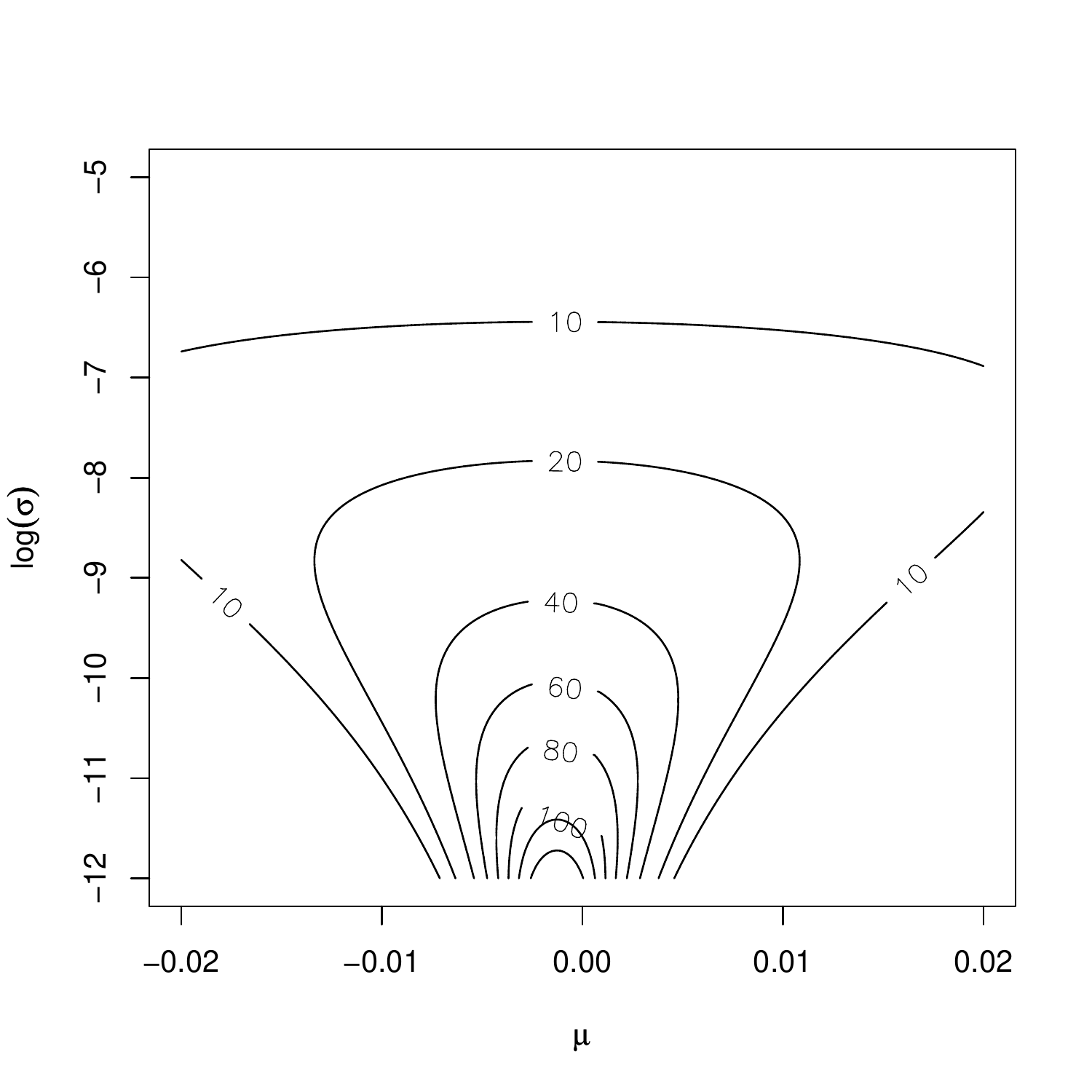}
\includegraphics[height=3.2in,angle=0]{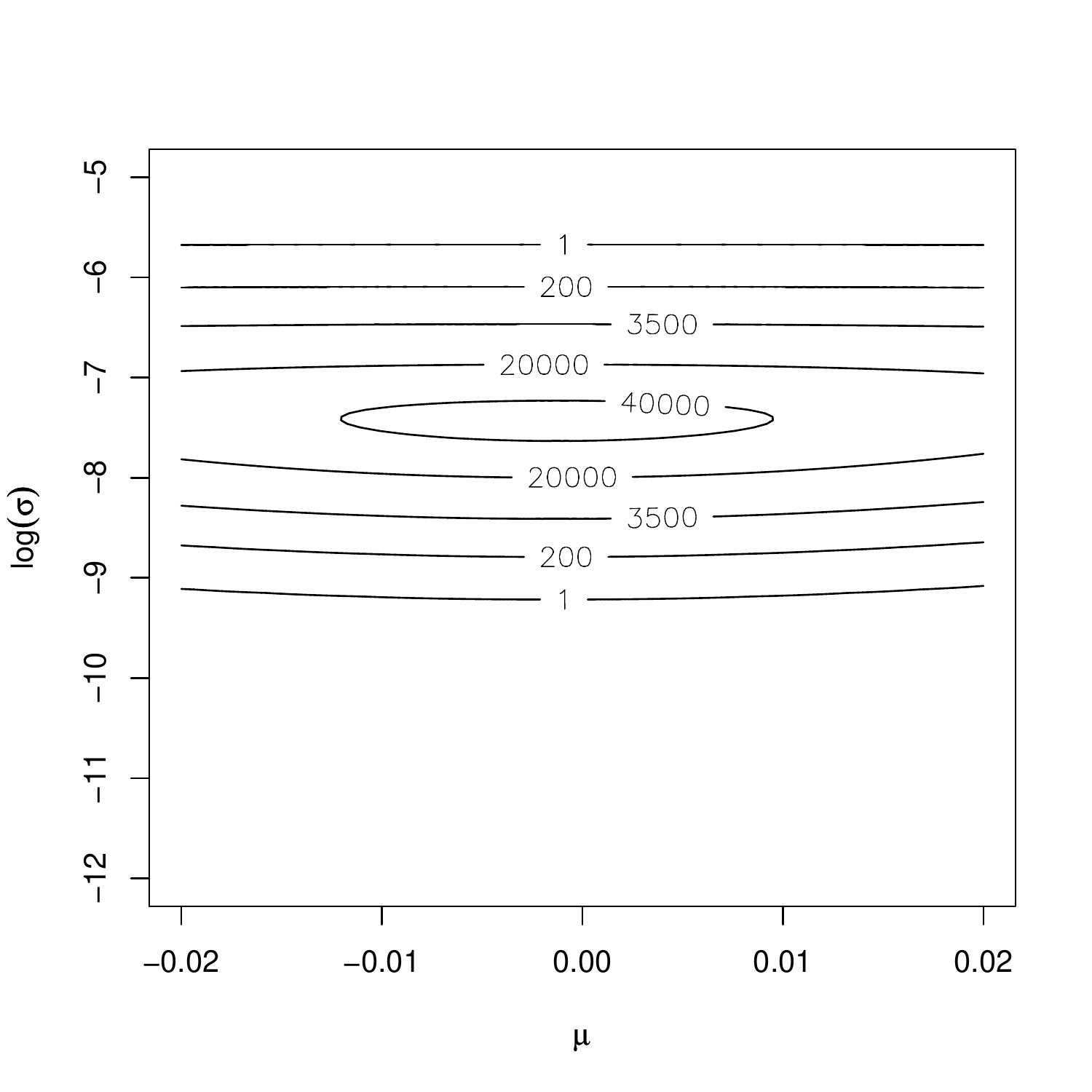} \\
\caption{Likelihood functions for a single observation. The left panel corresponds to the Gaussian likelihood obtained solely from the closing price (equation \eqref{eq:svdt_eq1}), while the right panel corresponds to the joint likelihood (equation \eqref{eq:jointpdf}).}\label{fi:likelihood}
\end{center}
\end{figure}

In order to emphasize the importance of the information provided by the minimum and maximum returns, we present in Figure \ref{fi:likelihood} plots of the likelihood function for the first observation of the S\&P500 dataset discussed in Section \ref{se:illus} (the week ending on April 21, 1997).  When only the closing price is available, the likelihood provides information about the drift of the process, but not about the volatility (note that in this case the likelihood is unbounded in a neighborhood of $\sigma_t=0$, and therefore the maximum likelihood estimator does not exist).  Therefore, information about the volatility in this type of model is obtained solely through the evolution of prices, and is therefore strongly influenced by the underlying smoothing process.  In other words, the volatility parameters are only weakly identifiable, with the identifiability begin provided by the autoregressive prior in \eqref{eq:svdt_eq2}.  In practice, this means that formally comparing alternative models for the volatility is extremely difficult. However, when the maximum and minimum are included in the analysis, the likelihood for a single time period does provide information about the volatility in that period, which can greatly enhance our ability to infer and test volatility models.

Although dealing with an infinite sum can seem troublesome, our experience suggests that a small number of terms (less than 20) suffice to provide an accurate approximation.  In addition, we note that since the general term of the sum is strictly decreasing for all the likelihoods discussed above, it is easy to implement an adaptive scheme that stops adding terms once the change in the value is smaller than a given tolerance.

Note that equation \eqref{eq:jointcdf} assumes that the log closing price at period $t-1$, $y_{t-1}$ is the same as the log opening price at period $t$ (call this $x_t$).  However, in some cases these two prices can differ; for example, wars or unexpected news can happen during holidays whenever markets are closed provoking jumps that are not governed by the diffusion processes.  As with other stochastic volatility models, we can easily include this type of ``weekend effect'' by realizing that equations \eqref{eq:jointcdf} and \eqref{eq:jointpdf} use $y_{t-1}$ simply as the initial value of the GBM; therefore, they remain valid if we substitute $y_{t-1}$ by $x_t$.  

\subsection{Variance evolution and prior specification}

The previous discussion focused on the characteristics of the likelihood function for the discretized process {\it conditional} on its volatility.  The full specification of the model also requires that we define the evolution of the volatility in time.  For illustrative purposes, this paper focuses on the simple autoregressive process for the log volatility described in Section \ref{se:svmodels},
\begin{align}
\log(\sigma_{t}) | \log(\sigma_{t-1}) &\sim \normal \left(  \alpha + \phi [\log(\sigma_{t-1}) - \alpha]  , \tau^2 \right) & \log(\sigma_0) &\sim \normal \left( \alpha, \frac{\tau^2}{1-\phi^2} \right)
\end{align}

However, we would like to stress that more complex stochastic processes can easily be incorporated in this models without significantly increasing the computational complexity.  For example, stationary $AR(p)$ processes for the log volatility (which can potentially provide information on quasi-periodicities in the volatility process) can be introduced using the prior specification discussed by \cite{HuWe99}.  Similarly, including volatility jumps is straightforward using Markov switching models \citep{CaLo07}, as well as estimating the parameters of stochastic volatility models with several factors varying at different time scales \citep{FouHanMol09}.

The model is completed by introducing priors for the unknown structural parameters in the models.  Following standard practice, we assume that
\begin{align}\label{eq:priors}
\mu &\sim \normal( d_{\mu}, D_{\mu} ) & \alpha & \sim \normal( d_{\alpha}, D_{\alpha}) &  \phi & \sim \bet( q_{\phi}, r_{\phi} ) & \tau^2 & \sim \IGam( u_{\tau}, v_{\tau}).
\end{align}

This choice of priors ensures that all parameters have the right support; in particular, the beta prior for $\phi$ ensures that $0 \le \phi \le 1$ (remember our discussion in Section \ref{se:svmodels}).  The eight hyperparametes $d_{\mu}$, $D_{\mu}$, $d_{\alpha}$, $D_{\alpha}$, $q_{\phi}$, $r_{\phi}$, $u_{\tau}$ and $v_{\tau}$ have to be chosen according to the available prior information about the problem at hand.  For example, it is natural to choose $d_{\mu}$ to be close to the market risk-free rate, while choosing $d_{\alpha}$ close the logarithm of the long-term average volatility for the asset.  In this paper, we avoid noninformative priors because the sequential Monte Carlo algorithms we employ for model fitting (see Section \ref{se:inference} below) require proper priors in order to be implemented.

\section{Inference using particle filters}\label{se:inference}

The use of simulation algorithms to explore the posterior distribution of complex Bayesian hierarchical models has become popular in the last 20 years.  In particular, Markov chain Monte Carlo (MCMC) algorithms, which generate a sequence of dependent samples from the posterior distribution of interest, have become ubiquitous.  For inference in non-linear state space models, sequential Monte Carlo algorithms, and in particular particle filters, have become a standard tool.  Particle filters use a finite discrete approximations to represent the distribution of the state parameters at time $t$ given the observations up to $t$, which in our case reduces to $p(\sigma_{t} | \{ y_l, b_l, a_l \}_{l=1}^{t})$), and sequentially updates it to obtain $p(\sigma_{t+1} | \{ y_l, b_l, a_l \}_{l=1}^{t+1})$.  As with other Monte Carlo approaches to inference, the resulting samples can be used to obtain point and interval estimates, as well as to test hypothesis of interest.  However, unlike MCMC algorithms, there is no need to check for convergence of the algorithm, study its mixing properties, or devising proposal distributions.    \cite{DoFrGo01} provides an excellent introduction to sequential Monte Carlo Methods.

Most sequential Monte Carlo algorithms are unable to handle structural parameters that do not evolve in time, and assume them fixed and known in advance.  However, in our stochastic volatility model  structural parameters such as $\mu$, $\alpha$, $\phi$ and $\tau^2$ are unknown and need to be estimated from the data.  Therefore, in the sequel we concentrate on a version of the auxiliary particle filter \citep{PiSh99} developed by \cite{LiWe01} for exactly this purpose.  Their algorithm introduces an artificial Gaussian perturbation in the structural parameters and applies an auxiliary particle filter to the modified problem.  In order to correct for the information loss generated by the artificial perturbation, the authors introduce a shrunk kernel approximation constructed in such a way as to preserve the mean and covariance of the distribution of the structural parameters.  This kernel density approximation usually works well in practice, producing accurate reconstructions of the posterior distribution of both structural and state parameters while avoiding loss of information that plagues the self-organizing state space models in \cite{Ki98}.

Since the artificial evolution is assumed to follow a Gaussian distribution, it is typically necessary to transform the structural parameters so that the support of their distribution is the whole real line. Therefore, we describe our algorithm in terms of the transformed structural parameter $\eta = (\eta_1,\eta_2,\eta_3,\eta_4) = (\mu, \alpha, \logit(\phi), \log(\tau^2)) \in \reals^4$. After choosing a discount factor $0.5 < \epsilon<1$ (controlling both the size of the perturbation and the level of shrinkage in the density estimator) and generating a sample of $N$ particle from the prior distributions in \eqref{eq:priors}, the algorithm proceeds by repeating the following steps for $t=1,\ldots, T$
\begin{enumerate}
\item For each particle $j=1,\ldots, N$, identify prior point estimates $(z_{t+1}^{(j)},m^{j}_{t+1})$ for the joint vector of state and structural parameters $(\sigma_{t+1}^{(j)}, \eta_{t+1}^{(j)})$ such that,
\begin{align*}
z_{t+1}^{(j)} &= \exp \{  \eta_{2,t}^{(j)} + \expit(\eta_{3,t}^{(j)})[\log(\sigma_{t}^{(j)})- \eta_{2,t}^{(j)}] \} \\
m_{t+1}^{(j)} &= a \eta_{t}^{(j)} + (1-a) \bar{\eta}_t
\end{align*}
where $a = (3\epsilon-1)/(2\epsilon)$ and $\bar{\eta}_t = \sum_{j=1}^{N} \eta_{t}^{(j)}$.

\item Sample auxiliary indicators $\xi_t^{(1)}, \ldots, \xi_t^{(N)}$, each one also taking values in the set $\{1,\ldots, N\}$, so that
$$
\Pr(\xi_t^{(j)} = k) = p(y_{t+1},b_{t+1},a_{t+1} |y_{t}, \mu = m_{1,t+1}^{k}, \sigma = z_{t+1}^{k})
$$
where $p(y_{t+1},b_{t+1},a_{t+1} |y_{t}, \mu, \sigma_t^2)$ is given in \eqref{eq:jointpdf}.

\item Generate a set of new structural parameters $\eta^{j}_{t+1}$ by sampling
$$
\eta^{(j)}_{t+1} \sim \normal( m_{t+1}^{\xi_{t}^{(j)}}, (1-a^2) V_t)
$$
where $V_t = \sum_{j=1}^{N} (\eta_t^{(j)} - \bar{\eta}_t)'(\eta_t^{(j)} - \bar{\eta}_t)/N$.

\item Sample a value of the current state vector
$$
\log(\sigma_{t+1}^{(j)}) \sim \normal( \eta_{2,t+1}^{(j)} + \expit(\eta_{3,t+1}^{(j)})[\log(\sigma_{t}^{(\xi_{t}^{(j)})})- \eta_{2,t+1}^{(j)}] , \exp(\eta_{4,t+1}^{(j)})  )
$$

\item Resample the particles according to probabilities
$$
\omega_{t+1}^{(j)} \propto \frac{ p(y_{t+1},b_{t+1},a_{t+1} |y_{t}, \mu = \eta_{1,t+1}^{k}, \sigma = \sigma_{t+1}^{j}) }{ p(y_{t+1},b_{t+1},a_{t+1} |y_{t}, \mu = m_{1,t+1}^{j}, \sigma = z_{t+1}^{j}) }
$$
\end{enumerate}

Although particle filters are not iterative algorithms and therefore mixing and convergence are not issues, sequential Monte Carlo algorithms might suffer from particle impoverishment.  Particle impoverishment happens when the particle approximation at time $t$ differs significantly from the approximation at time $t+1$, leading to a small number of particles receiving most of the posterior weight; in the worst case, a single particle receives all the posterior weight.  A similar issue arises in importance sampling algorithms, where efficiency decreases as the distribution of the importance weights becomes less uniform.  In the sequential Monte Carlo literature it is common to use the effective sample size (ESS) to monitor particle impoverishment,
$$
ESS_t = \frac{N}{1 + \frac{\V(\omega_t)}{\left[ \E(\omega_t) \right]^2}}
$$
Values of the ESS close to $N$ point to well behaved samplers with little particle impoverishment, while small values of $N$ usually indicate uneven weights and particle representations that might be missing relevant regions of the parameter space.

\section{Missing data}\label{se:missing}

In some instances (for example, non-exchange traded assets), data on the observed extremes might be unavailable or unreliable for some trading periods.  When MCMC algorithms are used for inference, their structure can be easily exploited to deal with this type of situation by adding an additional sampling step in which the missing or unreliable data is imputed conditionally on the current value of the parameters.  However, this type of iterative procedure is not available in particle filter algorithms such as the one described in Section \ref{se:inference}.  This means that implementation under missing data requires that we compute the corresponding marginal distributions from the joint density derived from \eqref{eq:jointcdf}, which are to be used in steps (2) and (5) of the particle filter instead of \eqref{eq:jointcdf}.

As we discussed in Section \ref{se:extremesv}, the marginal density of the log closing price is simply a normal distribution with mean $\mu$ and variance $\sigma_t^2$.  Therefore, if both the maximum and minimum are missing, the likelihood for the period simply becomes the likelihood for a standard stochastic volatility model in \eqref{eq:svdt_eq1}. When the minimum is not available, the marginal density for the maximum and the closing price is simply given by
\begin{eqnarray*}
q(y_{t},b_{t}|y_{t-1}) & = & \frac{2(2b_{t}-(y_{t}+y_{t-1}))}{\sqrt{2\pi\sigma_{t}^{6}t^3}}\exp\left\{-\frac{(2b_{t}-(y_{t}+y_{t-1}))^2}{2\sigma_{t}^{2}t}+\frac{\mu(y_{t}-y_{t-1})}{\sigma_{t}^{2}}-\frac{\mu^{2}t}{2\sigma_{t}^{2}}\right\},
\end{eqnarray*}
while if the maximum is missing, the marginal density for the minimum and the closing is given by,
\begin{eqnarray*}
q(y_{t},a_{t}|y_{t-1}) & = & \frac{2(y_{t}+y_{t-1}-2a_{t})}{\sqrt{2\pi\sigma_{t}^{6}t^3}}\exp\left\{-\frac{(y_{t}+y_{t-1}-2a_{t})^2}{2\sigma_{t}^{2}t}+\frac{\mu(y_{t}-y_{t-1})}{\sigma_{t}^{2}}-\frac{\mu^{2}t}{2\sigma_{t}^{2}}\right\}
\end{eqnarray*}
 The density of the maximum and minimum can be computed as well by integrating equation \eqref{eq:jointpdf} from $a$ to $b$ with respect to the variable $y_{t}$.  Proof of these results can be seen \citet{DanJean07}, or can be directly obtained by computing the corresponding limits on expression \eqref{eq:jointpdf}.


\section{Illustrations}\label{se:illusa}

\subsection{Simulation study}\label{se:illus}

In this section we use a simulation study to compare the performance of four stochastic volatility models:  STSV, which uses the Gaussian likelihood in \eqref{eq:constvar} and therefore employs solely the information contained in the opening and closing prices; RASV, which uses the likelihood  \eqref{eq:rangeonlypdf} and represents a Bayesian version of the range-only model described by \cite{AliBraDieb02}; RCSV, which uses the likelihood in \eqref{eq:rangeclosepdf} and therefore corresponds to the model based on the range and the opening/closing prices described in \cite{BraJon05}; and EXSV, our proposed model using the full likelihood \eqref{eq:jointpdf} and employing all the information contained in the opening/closing/high/low prices.

Our simulation study uses 100 random samples, each comprising 156 periods of returns generated under the stochastic volatility model described in equations \eqref{eq:svdt_eq1} and \eqref{eq:svdt_eq2}.  First, we generate the sequence of volatilities using the Ornstein-Uhlenbeck model in \eqref{eq:svdt_eq2}, with parameters $\alpha = -3.75$, $\phi = 0.9$ and $\tau = 0.11$ (note that these values correspond to the means of the prior distributions used for the analysis of the weekly S\&P500 data in Section \ref{se:sp500} below).  Then, for each period, we generate a sample path for the geometric Brownian motion in \eqref{eq:svdt_eq1} over a grid with 1000 nodes using the corresponding value for the volatility, assuming that $\mu=0.000961$ (for weekly returns, this corresponds approximately to an average 5\% annual return).  The maximum, minimum, opening and closing prices over the period were computed from this sample path.  We assume that the value of the assets at the beginning of each simulation $S_0$ is \$100 and compute both the root mean squared deviation (RMSD) of the estimated volatility with respect to their true value, and their median absolute deviation (MAD).

The particle filter algorithm described in Section \ref{se:inference} was used to fit all four models, and was implemented in {\tt MATLAB} using 30,000 particles.  Unless noted otherwise, the results we report below correspond to prior hyperparameters $d_{\mu} = 0$, $D_{\mu} = 0.0001$, $q_{\phi}=9$ and $r_{\phi}=1$, $d_{\alpha} = -3.75$, $D_{\alpha} = 0.025$ and $u_{\tau} = 6$ $v_{\tau} = 0.06$, so that the prior distributions are centered around the true parameter values for all models.

Table \ref{ta:simstudy} compares the performance of the different stochastic volatility models in this simulation scenario. The first number in each cell corresponds to the median ratio of the pertinent performance measure across 100 simulations, while the numbers in parenthesis correspond to the 5\% and 95\% quantiles, respectively.  Ratios close to one indicate that the two models in question have similar performance, while ratios much larger than one indicate that the first model has a much larger RMSD (or MAD) than the second one, and therefore performs worse.
\begin{table}
\begin{tabular}{|c|c|c|} \hline
Models & RMSD & MAD \\ \hline\hline
STSV/RASV  & 1.43 (1.21, 1.64) & 1.46 (1.05, 1.82) \\ 
RASV/RCSV  & 1.02 (0.90, 1.13) & 0.99 (0.85, 1.16) \\ 
RCSV/EXSV  & 1.06 (0.92, 1.22) & 1.11 (0.96, 1.26) \\
RASV/EXSV  & 1.07 (0.97, 1.18) & 1.10 (0.93, 1.27) \\ \hline
\end{tabular}
\caption{Results from our simulation example.  We show the median along with the 5\% and 95\% quantiles (in parenthesis) of the ratio between deviation measures for three pairs of models, computed over a total of 100 simulated data sets.}\label{ta:simstudy}
\end{table}

In agreement with the results reported by \cite{AliBraDieb02}, the first row of Table \ref{ta:simstudy} shows that using the range alone as a volatility proxy produces consistently more accurate volatility estimates than using opening and closing prices alone.  Similarly, the second row suggests that, although including the opening prices tends to improve the volatility estimates over those obtained from the range-based model most of the time, it can sometimes decrease accuracy (specially when the MAD is used to measure the accuracy of the reconstruction).  A similar situation, although much less severe, happens with our model based on the full likelihood;  our model tends to improve over RASV and RCSV most of the time (with an average efficiency gain in the range of 5-10\%), but can underperform in some data sets.

Although the impact of the full information on the estimation of the volatility is moderate, including the full information contained in CHLO data can greatly improve the estimation of the drift of the model.  For example, the ratio of the absolute error in the estimate of the drift $\mu$ based on the 156 time points between RASV and EXSV has a median of 1.12, with the 5\% and 95\% quantiles being 0.93 and 1.46.  This indicates an average gain in efficiency of about 12\%.  We also note that EXSV tends to be more robust to prior misspecification.  Sensitivity analysis performed using different hyperparameters (results not shown) showed that estimates tend to be less affected by prior choice under our EXSV model.

\begin{figure}
\begin{center}
\includegraphics[height=3.0in,angle=0]{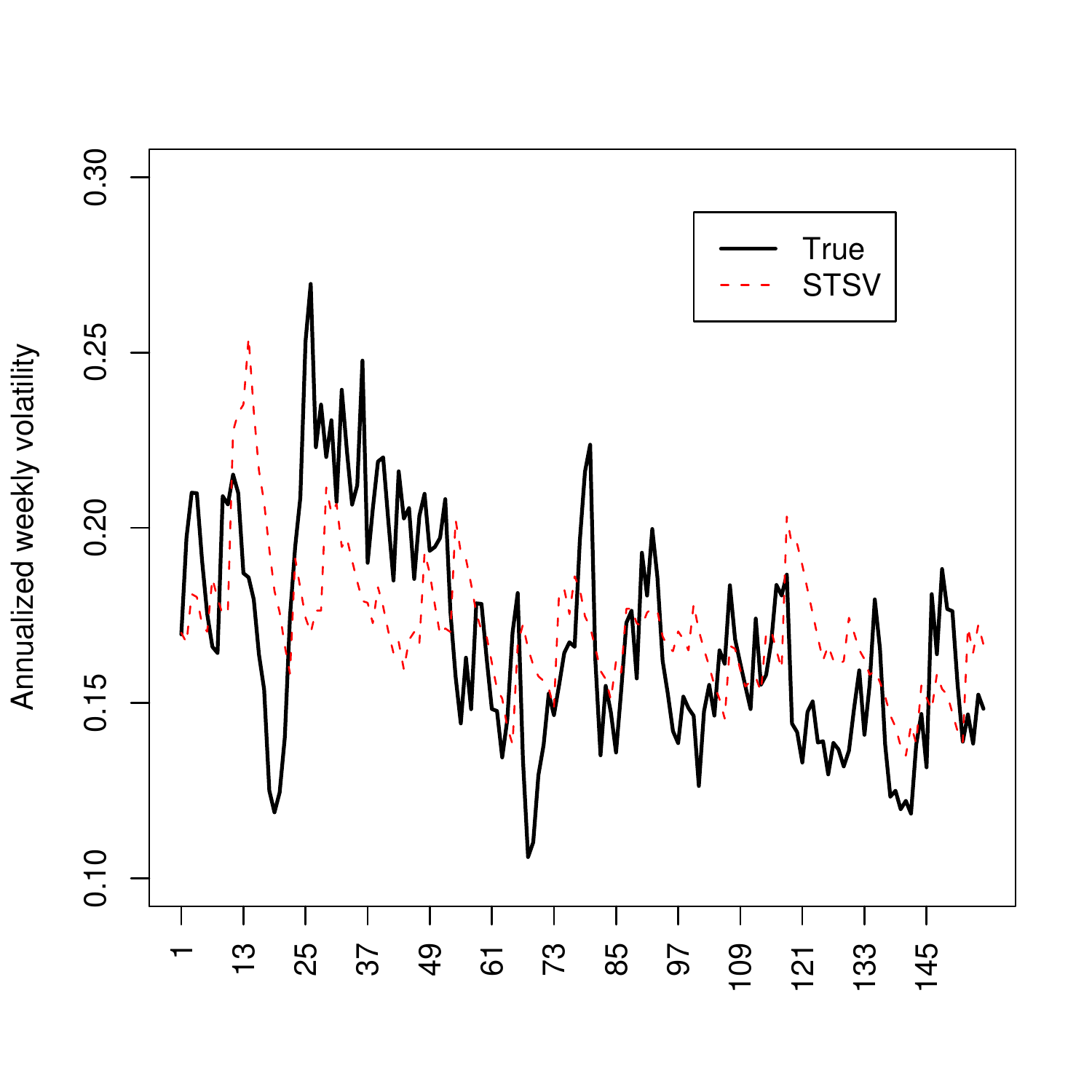}
\includegraphics[height=3.0in,angle=0]{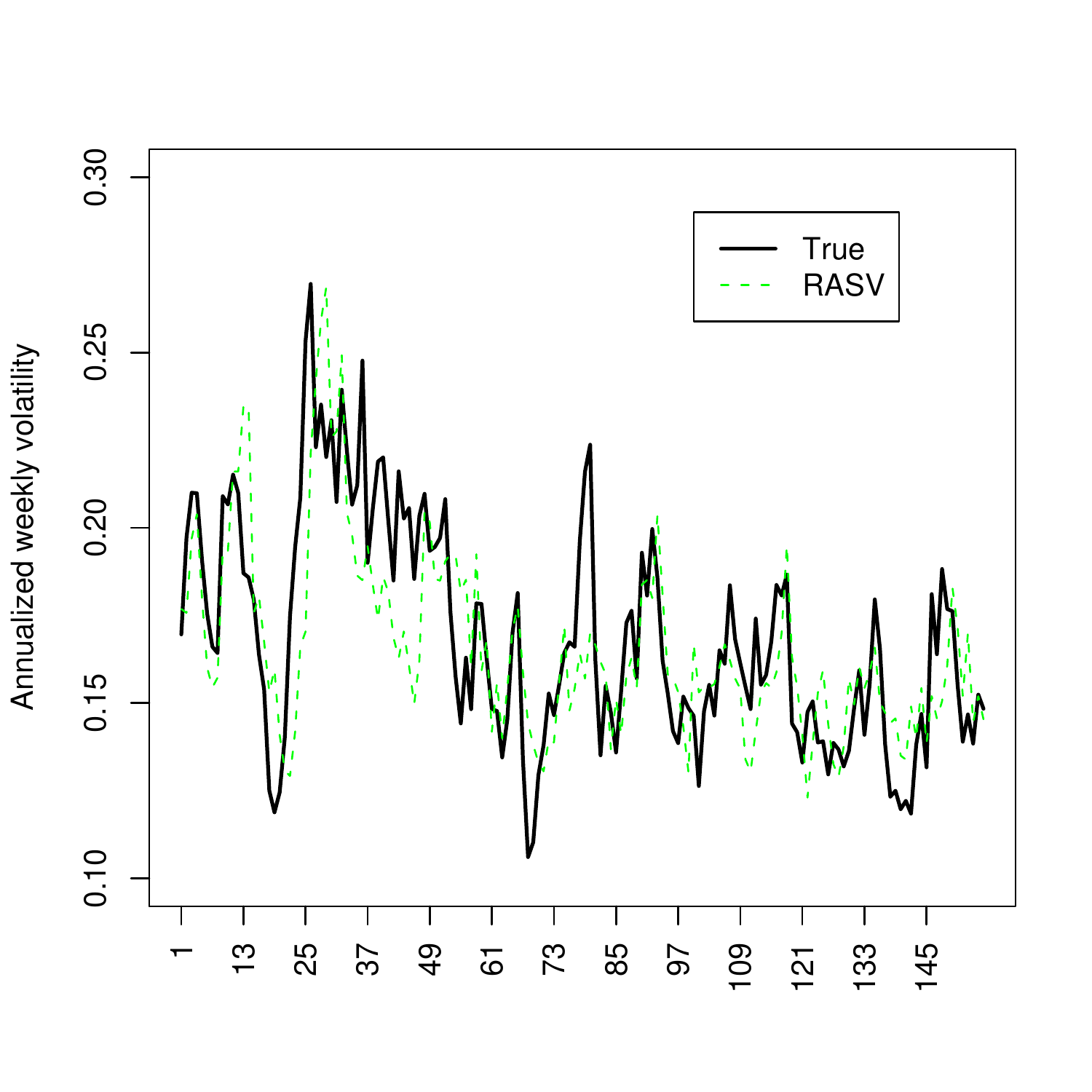}\\
\includegraphics[height=3.0in,angle=0]{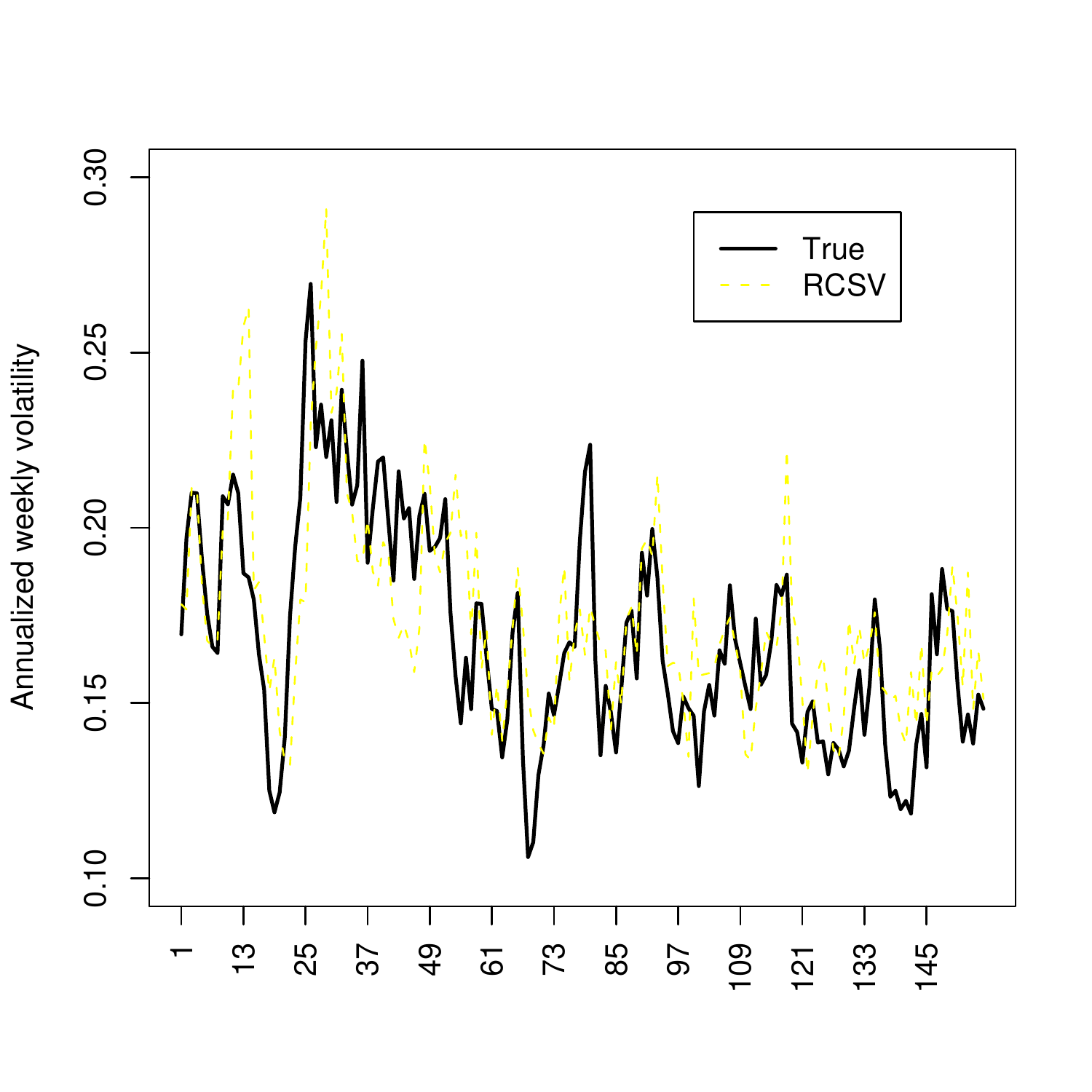}
\includegraphics[height=3.0in,angle=0]{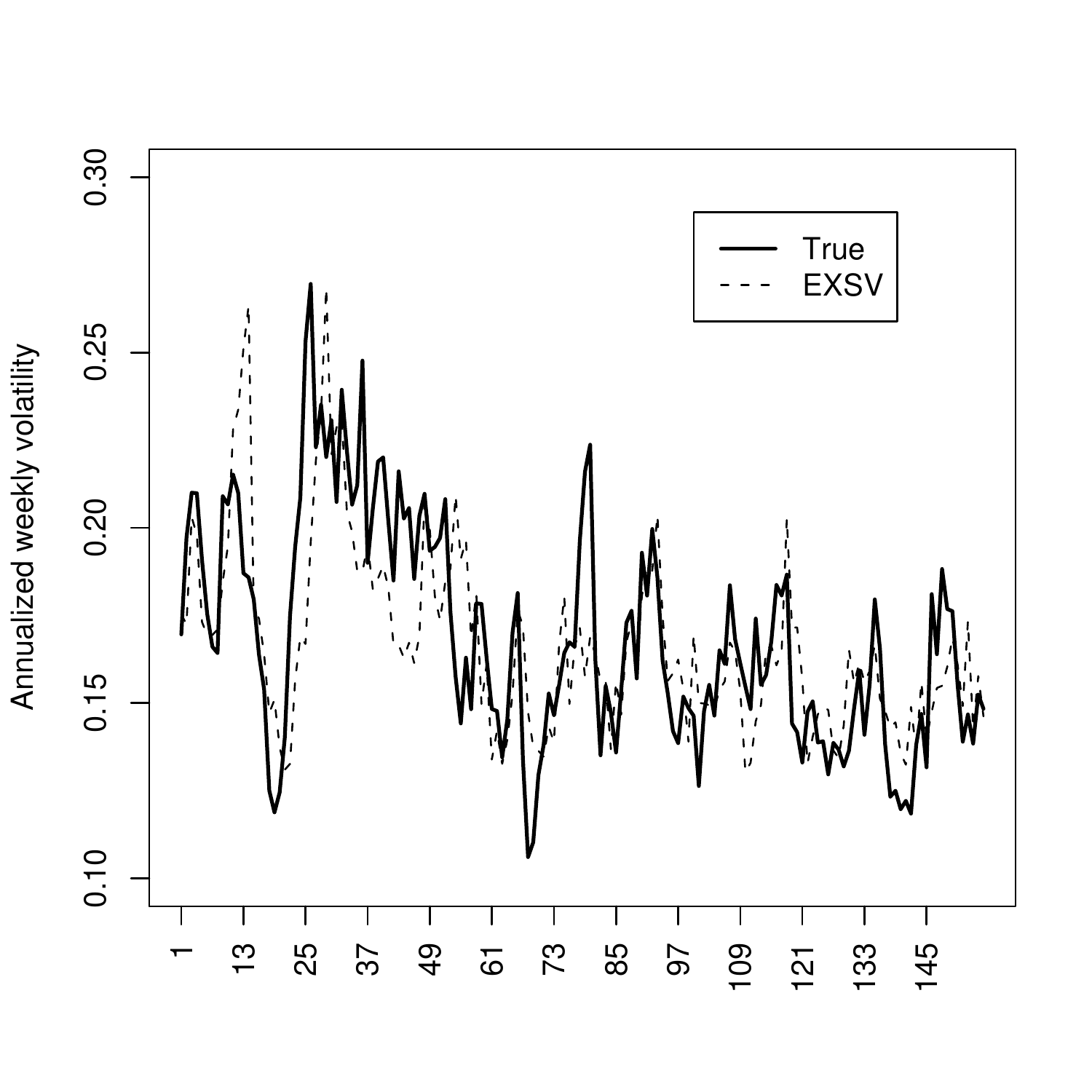}
\caption{True and reconstructed volatility paths for one simulation in our study for STSV (top left panel), RASV (top right), RCSV (bottom left) and EXSV (bottom right).}\label{fi:rsimstdy}
\end{center}
\end{figure}

Further insight into the behavior of these four models is provided by Figure \ref{fi:rsimstdy}, which shows the true and reconstructed volatility paths for one simulation in our study.  Note that paths generated by STSV tend to underestimate the volatility of volatility.  Also, the paths from RASV, RCSV and EXSV tend to be quite similar to each other, specially RCSV and EXSV.  These results suggest that much of the information about the volatility contained in the High and Low prices is provided by the range, but also that the actual levels of the minimum and maximum can provide additional helpful information in most cases.

\subsection{Estimating the volatility in the S\&P500 index}\label{se:sp500}

\begin{figure}
\begin{center}
\includegraphics[height=3.2in,angle=0]{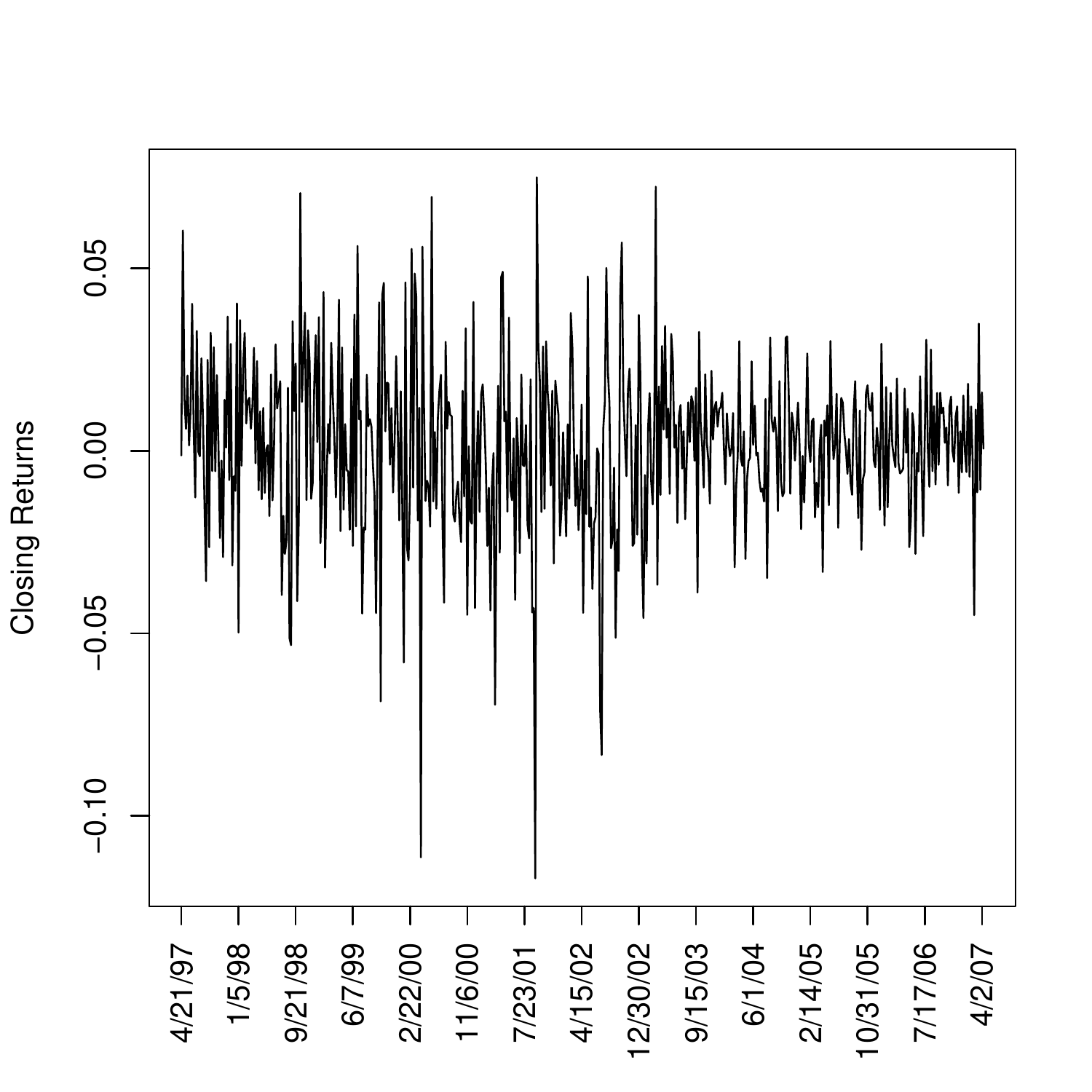}
\includegraphics[height=3.2in,angle=0]{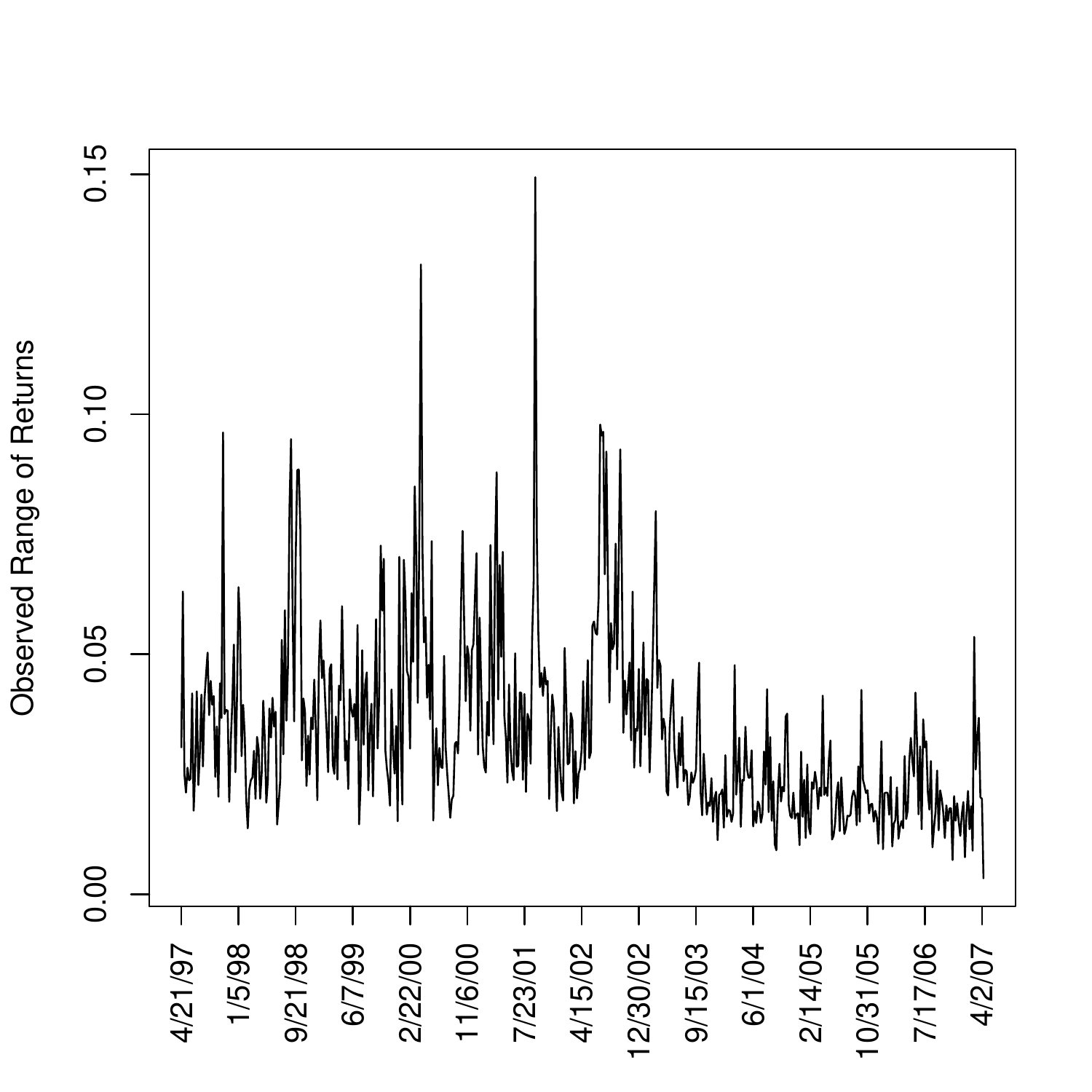} \\
\caption{Observed weekly closing returns (left panel) and return ranges (right panel) in the S\&P500 data.}\label{fi:rawreturns}
\end{center}
\end{figure}

In this section we consider the series of the weekly S\&P500 prices covering the ten-year period between April 21, 1997 and April 9, 2007, for a total of 520 observations.  Figure \ref{fi:rawreturns} shows the evolution of the log returns at closing, as well as the observed ranges in log returns (computed as the maximum observed log return minus the minimum observed log return over the week).  Note that both plots provide complementary but distinct information about the volatility in prices.  The series of closing returns does not exhibit any long term trend, but different levels of volatility can be clearly seen from both plots, with the period 1997-2002 presenting a higher average volatility than the period 2003-2007.

The data was analyzed using three of the models considered in the previous section:  STSV, RASV and EXSV.  As in the previous section, prior hyperparameters were chosen so that $d_{\mu} = 0$, $D_{\mu} = 0.0001$ (therefore, we expect the average weekly returns to be between -0.03 and 0.03 with high probability), $q_{\phi}=9$ and $r_{\phi}=1$ (so that we expect the autoregressive coefficient to be around $0.9$), $d_{\alpha} = -3.75$, $D_{\alpha} = 0.025$ and $u_{\tau} = 6$ $v_{\tau} = 0.06$ (so that the median of the annualized long-term volatility is a priori around $20\%$).  The models were fitted using the particle filter algorithm described in Section \ref{se:inference}.  We used a total of 100,000 particles for each model, and monitored particle impoverishment by computing the effective sample sizes at each point in time.  Although not a serious issue in any of the three models, particle impoverishment is more pronounced for EXSV.  This is probably due to the additional information provided by the extremes, which makes the likelihood tighter and decreases the viability of individual particles.
\begin{figure}
\begin{center}
\includegraphics[height=3.2in,angle=0]{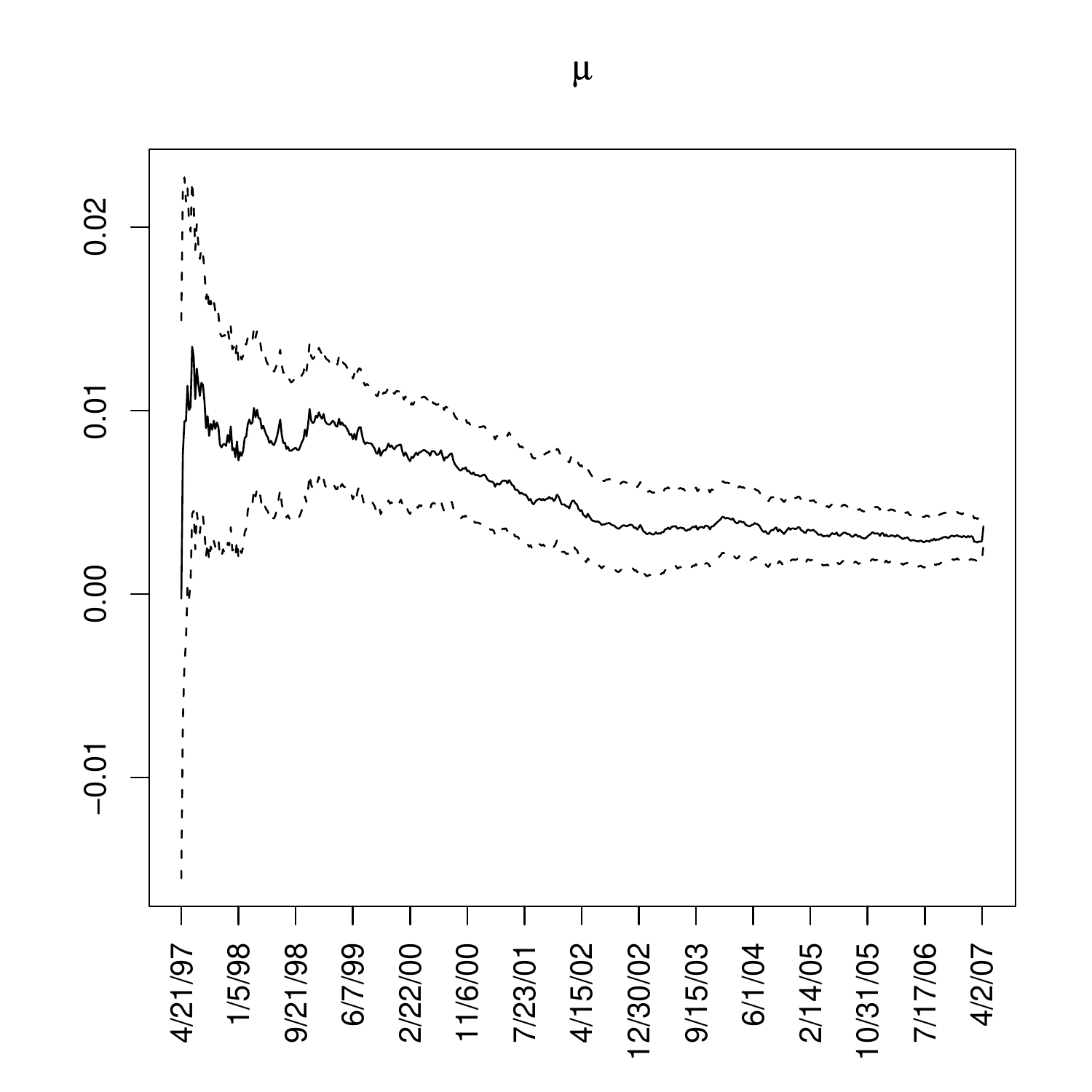}
\includegraphics[height=3.2in,angle=0]{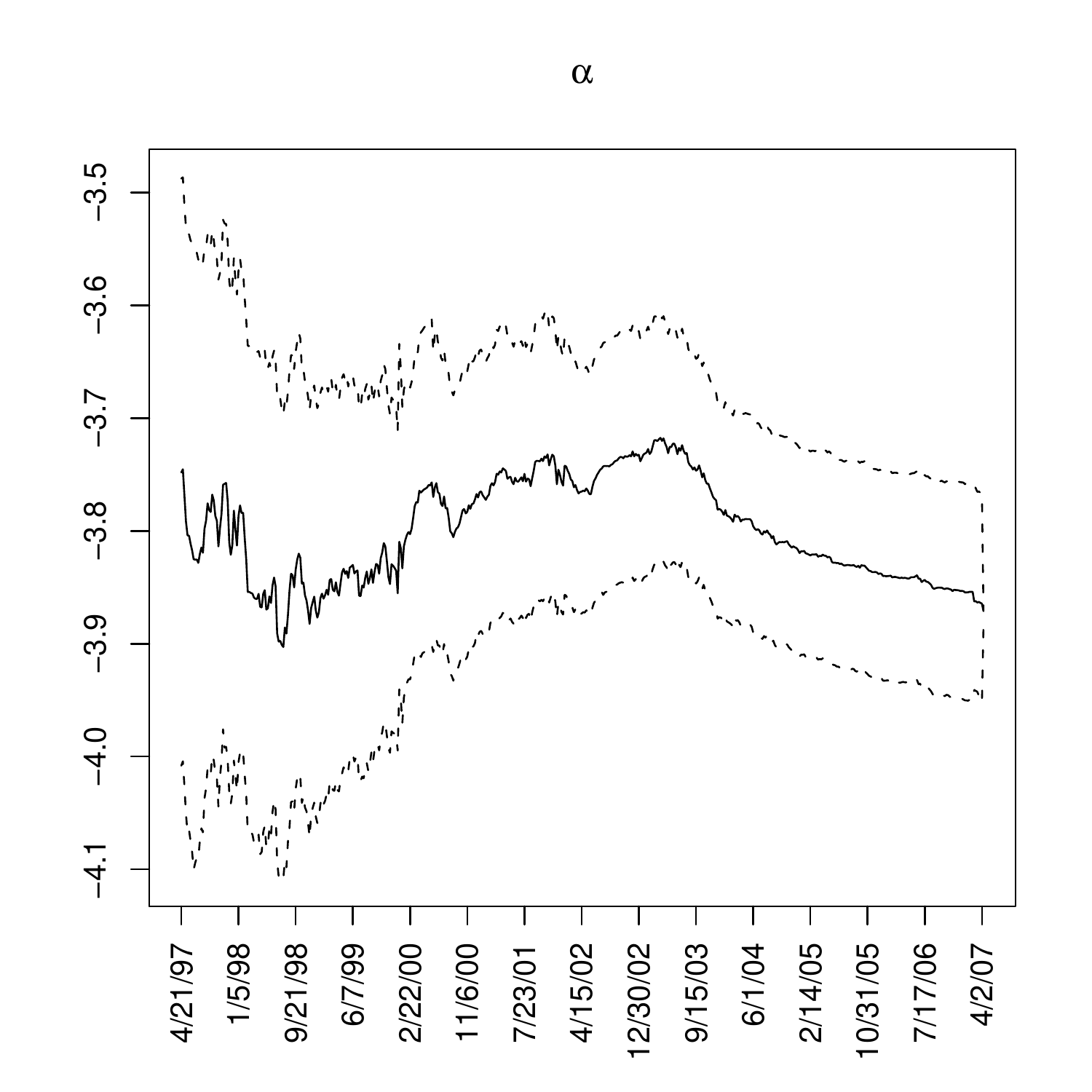} \\
\includegraphics[height=3.2in,angle=0]{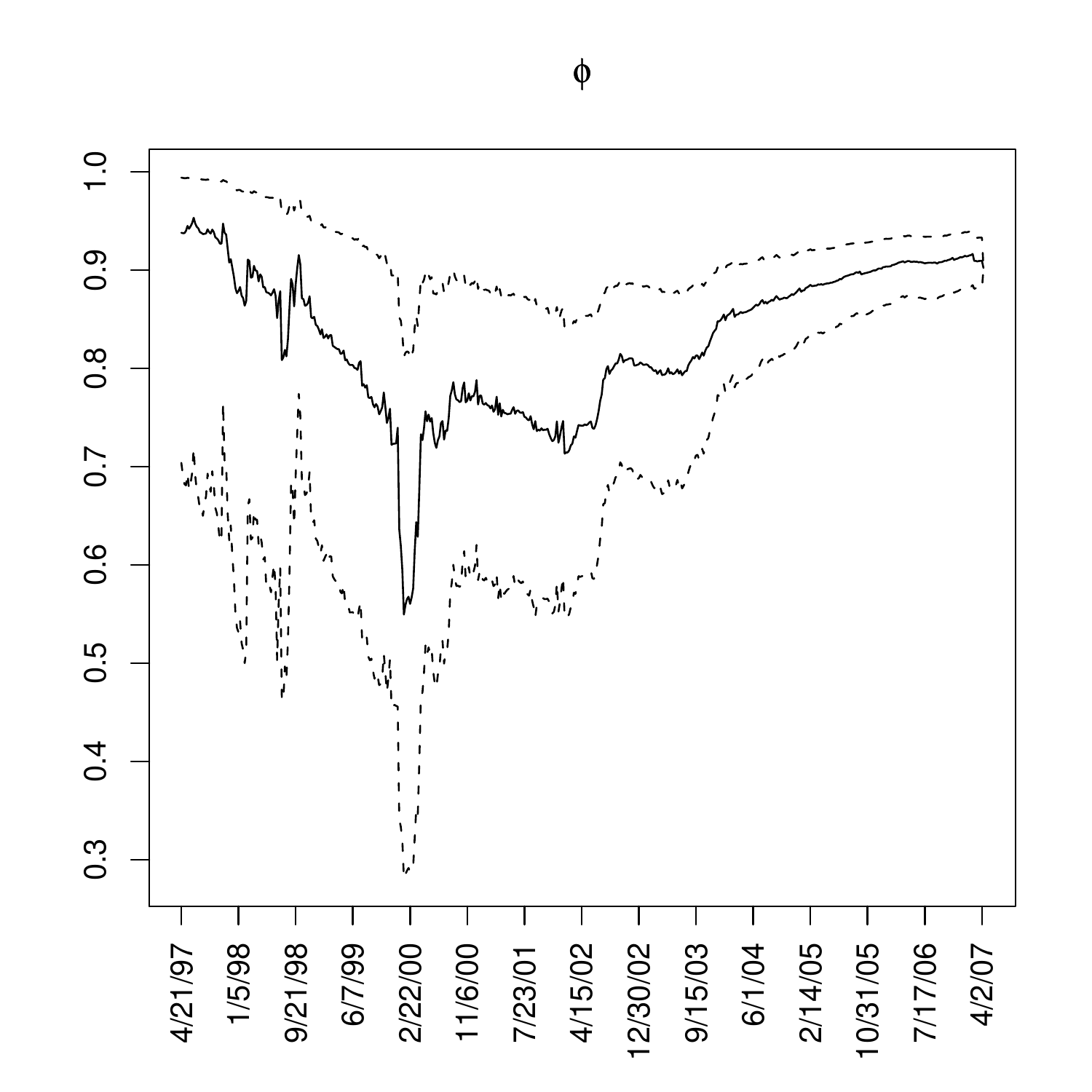}
\includegraphics[height=3.2in,angle=0]{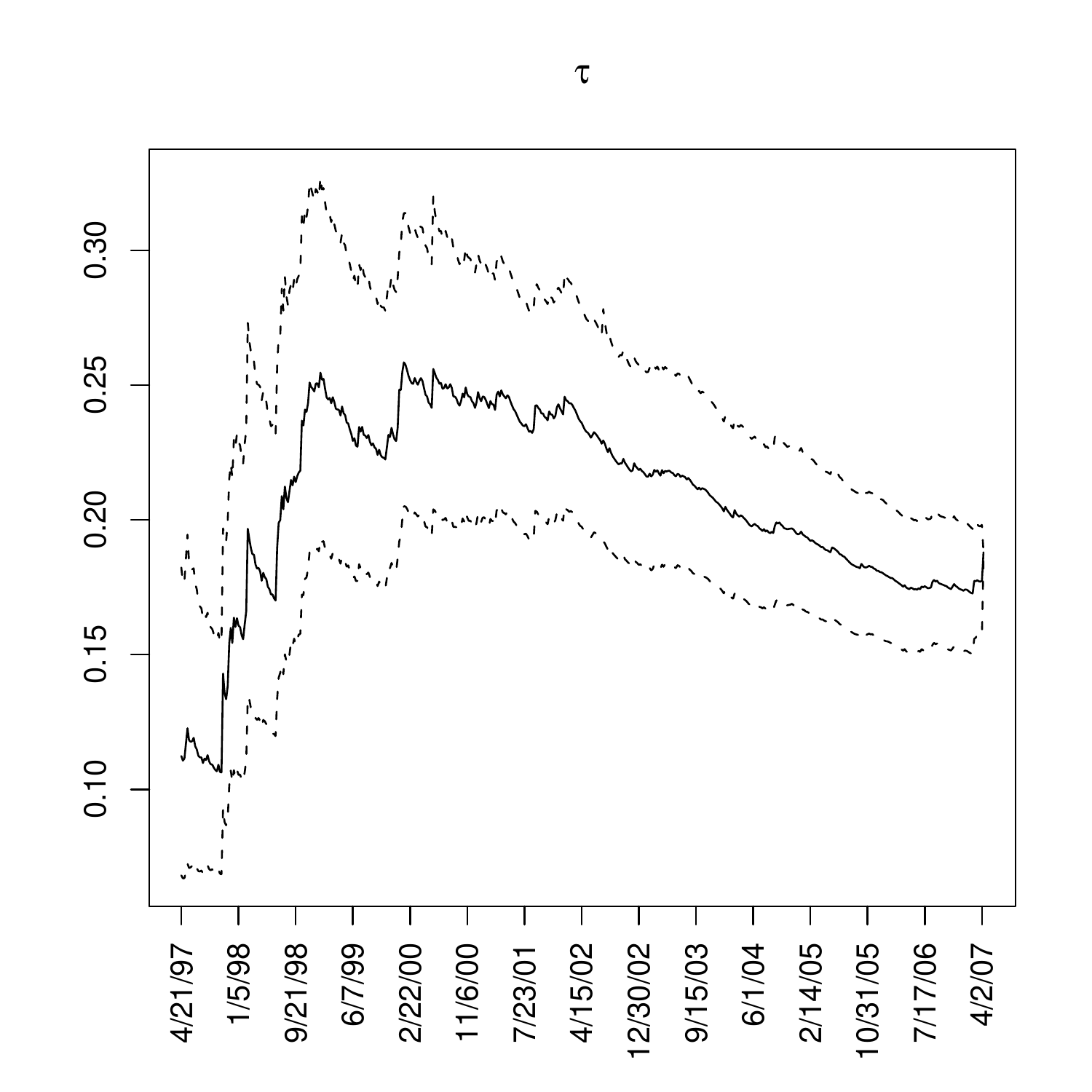} \\
\caption{Filtered means and 5\% and 95\% posterior quantiles for the structural parameters under EXSV.}\label{fi:structural}
\end{center}
\end{figure}

Point and interval estimates for the structural parameters under EXSV are shown in Figure \ref{fi:structural}.  These are filtered estimates, which means that they incorporate information available only until the time they were computed.  We note that there is substantial learning about the structural parameters.  After $T=520$ observations, the posterior mean for the median of the stationary distribution of volatility, $\nu = \exp\{ \alpha \}$, is $0.1546$, with a symmetric 90\% credible interval $(0.1510, 0.1670)$, while the autoregressive coefficient for the volatility has a posterior mean of $0.8935$ with credible interval $(0.8832, 0.9134)$.  The results from STSV (not shown) are similar, but tend to produce a larger value of the autocorrelation coefficient (posterior mean 0.9690, 90\% credible interval (0.9245, 0.9873)).  One surprising feature of these estimates is the pronounced drop in the autocorrelation coefficient $\phi$ on the week of January 24, 2000, which is accompanied by an increase in the volatility of the volatility, $\tau$.  This can be explained by the large negative return observed for this week (around -11\% in the week).  This observation suggests that a model that includes volatility jumps would be more appropriate for this data, however, the development of such a model is beyond the scope of this paper and will be discussed elsewhere.
\begin{figure}
\begin{center}
\includegraphics[height=4.9in,angle=0]{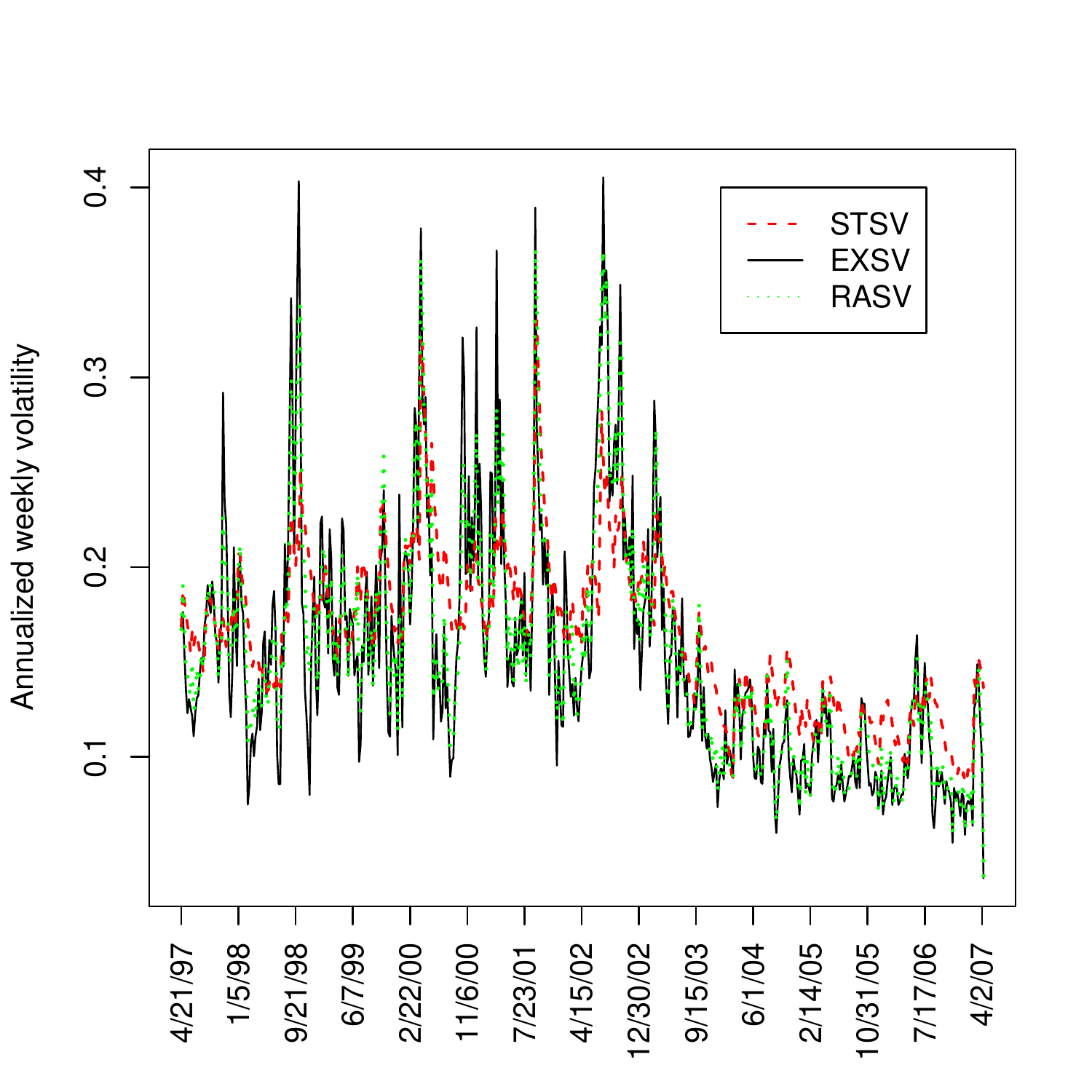}
\caption{Estimated volatilities of S\&P500 data using both stochastic volatility models.}\label{fi:filtered}
\end{center}
\end{figure}

\begin{figure}
\begin{center}
\includegraphics[height=3.2in,angle=0]{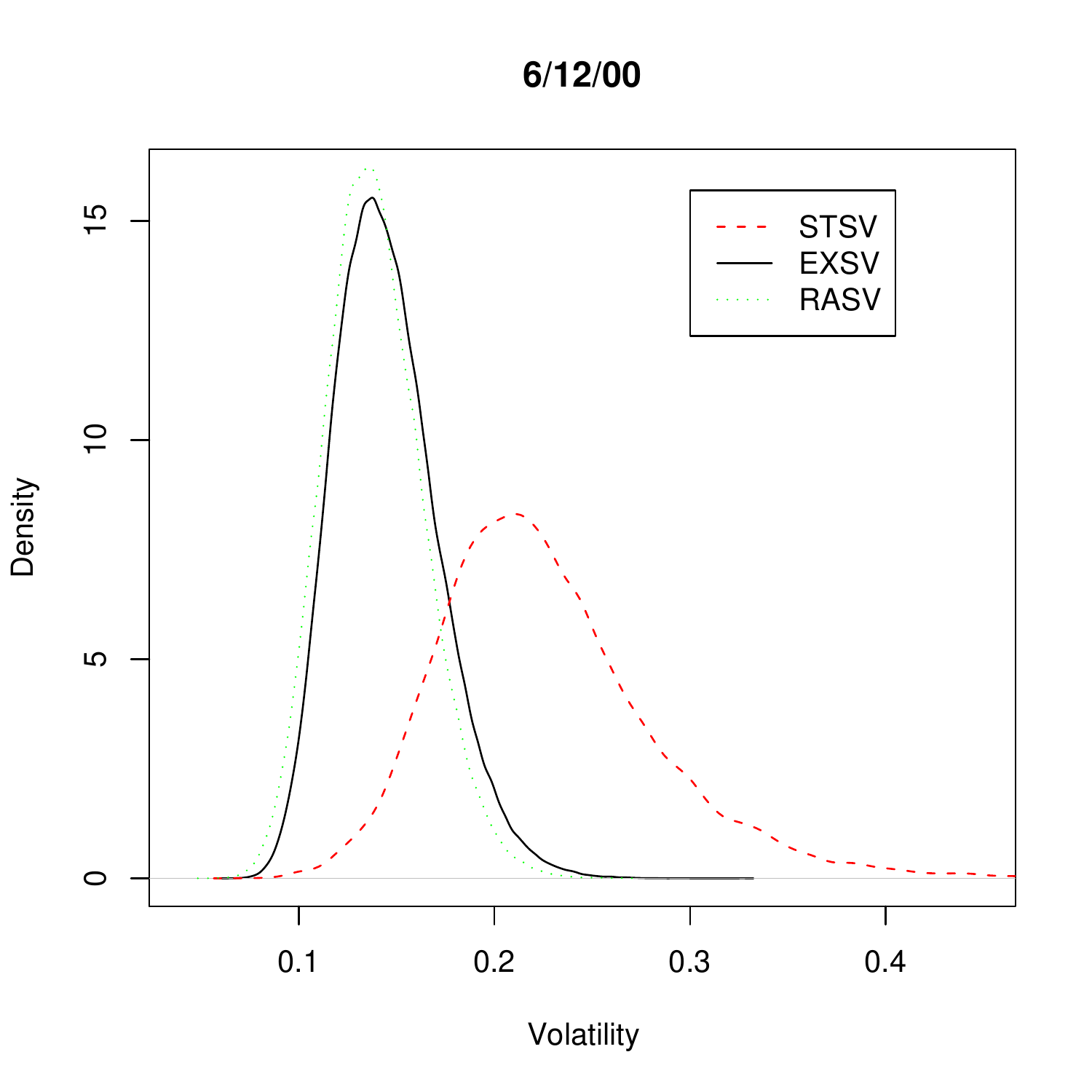}
\includegraphics[height=3.2in,angle=0]{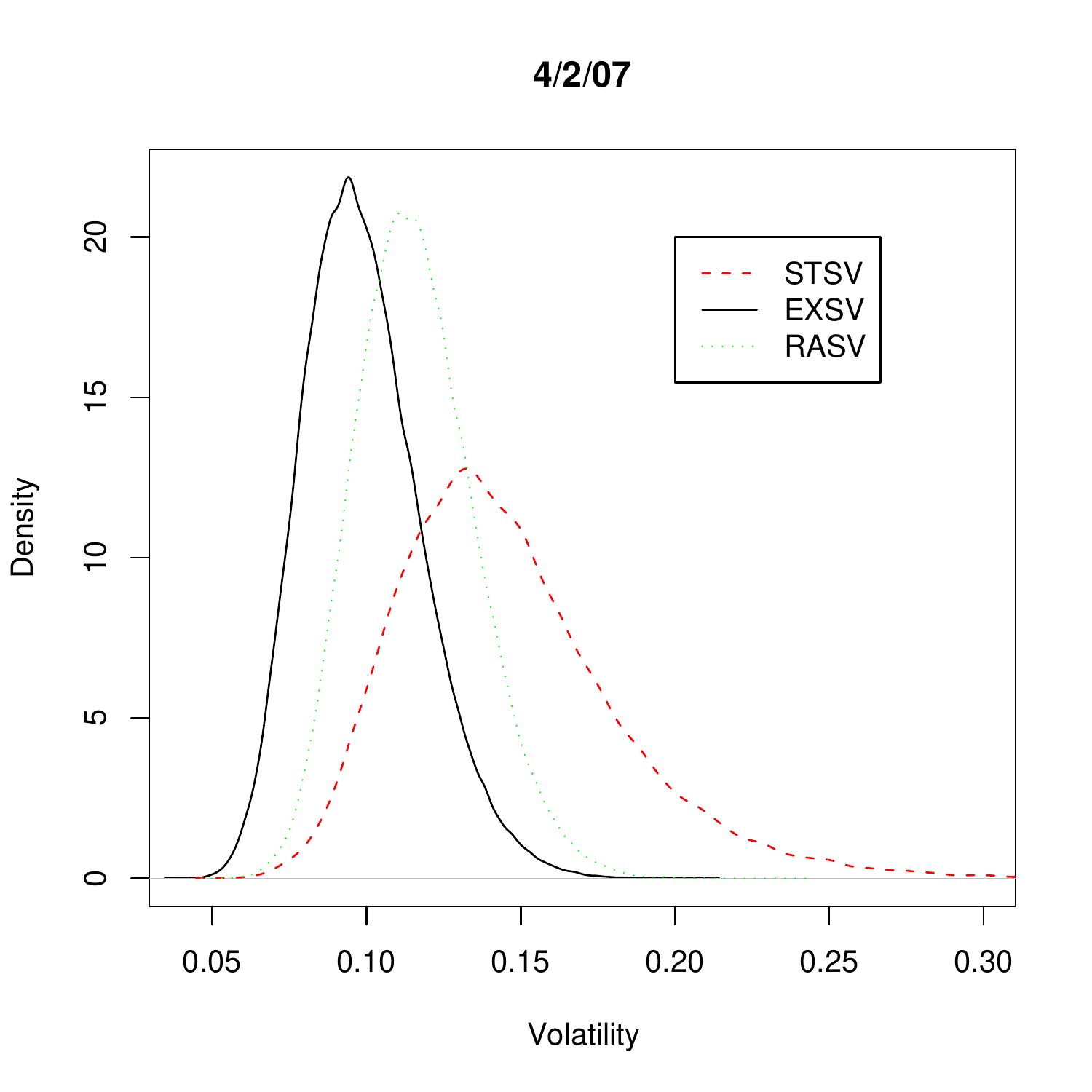}
\caption{Posterior distribution for the volatility of returns for December 6, 2000 (left panel) and April 2, 2007 (right panel)}\label{fi:postdenest}
\end{center}
\end{figure}

Figure \ref{fi:filtered} shows point estimates for the volatility of returns under all three models. Again, these are filtered estimates.  Although the overall level of the volatility series seems to be similar in all three instances, and the estimates obtained from RASV and EXSV are similar, there are striking differences in the behavior of STSV with respect to RASV and EXSV.  For example, note the peaks during the high volatility period of 1997-2002 are much more pronounced under RASV and EXSV, while the average volatility between 2003 and 2007 seems to be lower under both RASV and EXSV than under STSV.  Therefore, including information on extreme values seems to help to correct not only for underestimation, but also for overestimation of the volatility.  To complement the information in Figure \ref{fi:filtered}, we present in Figure \ref{fi:postdenest} the posterior distribution for the volatility of returns on December 6, 2000 and April 2, 2007 under all three models.  All distributions are right skewed, but the distributions under RASV and EXSV are shifted to the left, which suggest that STSV overestimates the true volatility of model in both cases.  In addition, for December 6, 2000 the distributions for RASV and EXSV are almost identical; however, for April 2, 2007, they are noticeable different.  More important, the posterior under both EXSV and RASV has a lower variability than that under STSV, reflecting the additional information contained in the extreme values.

\begin{figure}
\begin{center}
\includegraphics[height=3.0in,angle=0]{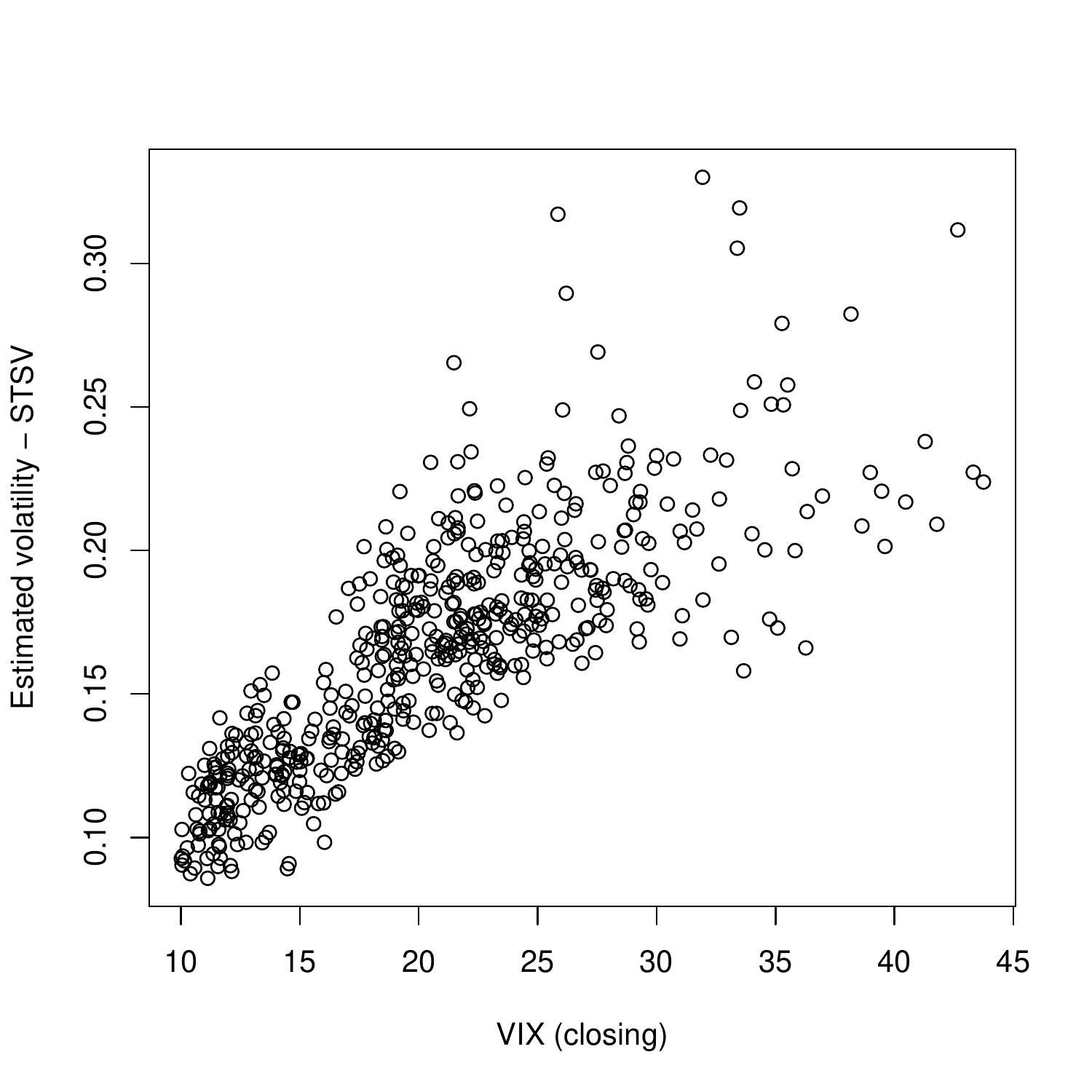}
\includegraphics[height=3.0in,angle=0]{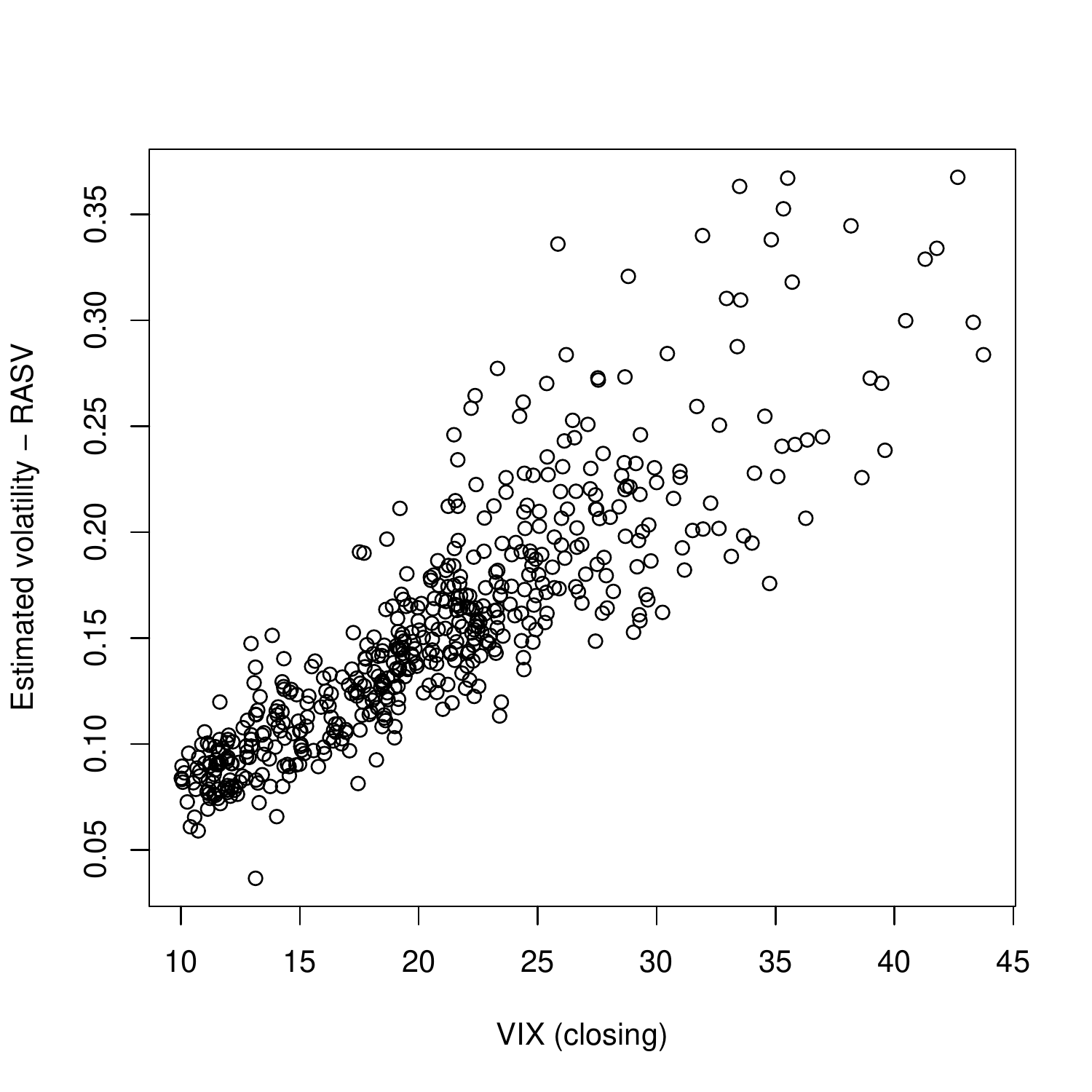}
\includegraphics[height=3.0in,angle=0]{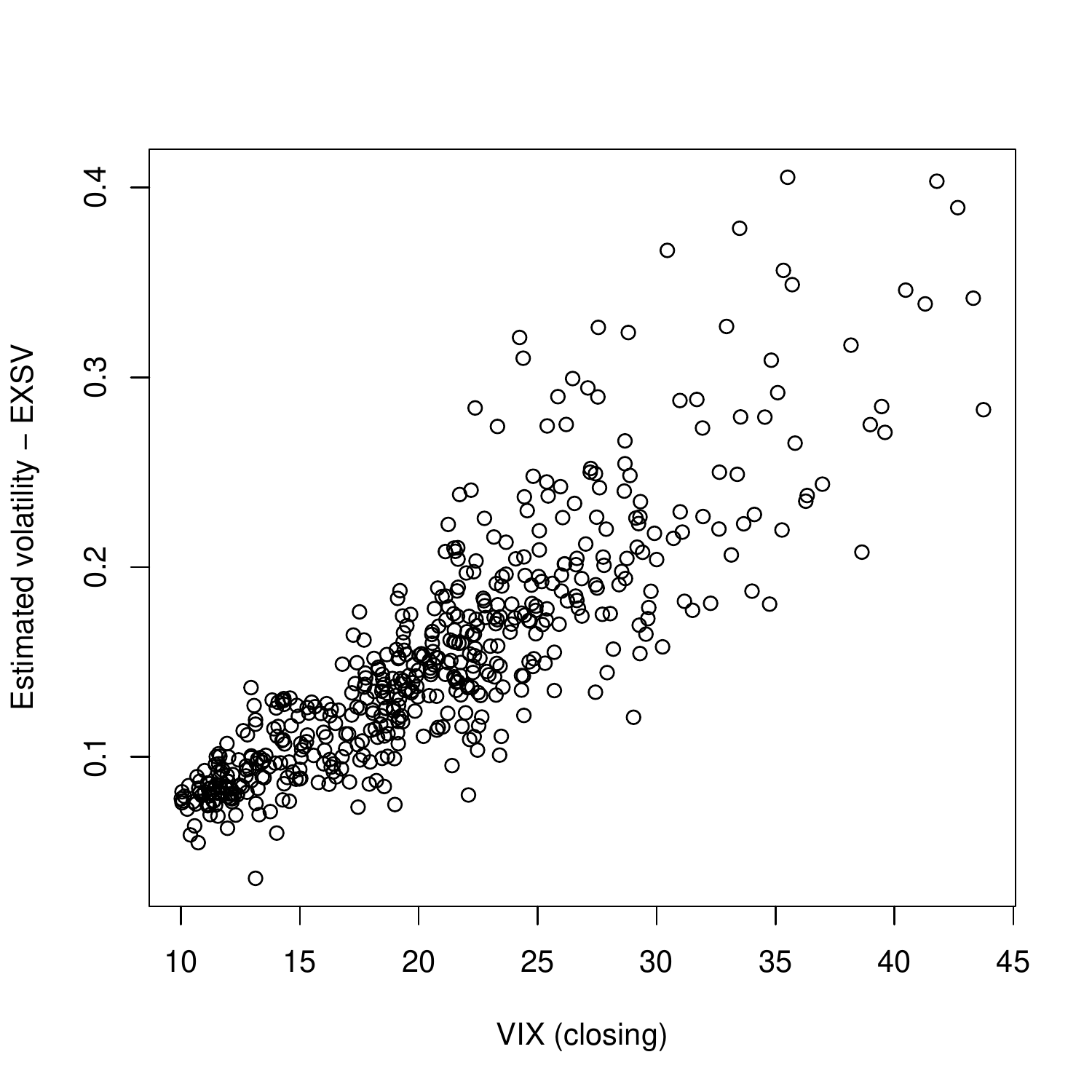} \\
\caption{Closing VIX prices versus filtered volatilities for all three models under consideration.}\label{fi:vix}
\end{center}
\end{figure}

In order to better understand the effect of the extreme values on the estimation of volatility, we compare our volatility estimates with the value of VIX index for the dates under consideration, which provides an independent source of information about the volatility on the S\&P500 index.  The VIX is constructed as a weighted average from the implied volatilities obtained from options whose underlying asset are included in the S\&P500, and are therefore not directly linked to the S\&P500 extremes.   Of course, the VIX is the implied forward looking volatility priced in the market, with 1-2 months average tenors and which includes the price of risk, while our estimates refer to current filtered estimates, but we make this comparison under the assumption that the VIX will reflect current market expectations of where the volatility of the underlying is, and any bias will affect equally all models under consideration.  Figure \ref{fi:vix} presents scatter plots of the observed VIX prices at closing, which show strong association with the volatilities estimated under all models.  Correlation between the closing VIX price and the filtered volatilities is 0.801 for the STSV model 0.853 for RASV and 0.861 for EXSV.  Very similar results are obtained if the average between the opening and closing VIX price are used instead (correlations are 0.816 for STSV, 0.854 for RASV and 0.863 for EXSV).  The stronger correlation with EXSV volatilities provides additional evidence that the information contained in the realized extremes can indeed improve the performance of the model, although the difference between RASV and EXSV is small and might not be relevant.

\section{Discussion and future work}\label{se:ccl}

This paper adds to the growing body of literature suggesting that the information contained in the observed extremes of asset prices can greatly contribute to increase the accuracy and efficiency volatility estimates.  In addition, our results also suggest that the information contained in the level of the extreme returns (which is lost when using the observed ranges for inferences) can also contribute to more efficient estimation of the volatility, and almost as important in certain applications, to the estimation of the drift of the process.  This paper also provides a unifying framework in which to understand several CHLO models available in the literature, allowing for a direct comparison of different assumptions.

Some extensions of the simple stochastic volatility model discussed in this paper are immediate and have already been hinted at.  For example, it is well known that leverage effects can be incorporated by introducing a correlation between the innovations in the price and volatility processes.  In addition, non-equally spaced observation can be easily accommodated by slightly rewriting the likelihood equations in terms of non-unit length intervals. The strong identifiability of model parameters provided by using the full information in the model allows us to reliably compare different evolution dynamics (e.g., Ornstein-Uhlenbeck, Markov-switching, jump and squared root processes).  Indeed, model comparisons in this setting is straightforward thanks to our reliance on particle filters for computation.  Finally, we plan to investigate how the additional information provided by the range can help when reconstructing option prices.  In this regard, note that, conditionally on the parameters of the underlying stochastic process, the pricing formulas in \cite{He93} can be used almost directly in our problem.  Therefore, our model can be extended to generate a posterior distribution for the price of any option of interest, which can be compared to the prices observed in the market.

Another avenue that deserves future exploration is the use of extreme prices for the estimation of covariance across multiple asset prices.  In that regard, we can start by writing a model for the multivariate log-asset prices $\bfy_t$ as a multivariate GBM, with time variance covariance matrix, and deriving the joint density for closing, high and low prices conditional on the opening prices, in a manner similar to what we did in this paper.  This joint likelihood contains all relevant information in the extremes and avoid the calculation of cross ranges \citep{RoZh08}, which scale badly to higher dimensions.
\appendix

\section{Proof of theorem 1}\label{ap:th1}

Consider $X_t$, a driftless Brownian motion with unit variance starting at zero.  In Chapter 3, p. 75 of \cite{Kl05}, the joint cumulative distribution for the maximum $M_t = \max_{0 \le s \le t}\{ X_s \}$, minimum $m_t =\min_{0 \le s \le t} \{  X_s \}$ and final value of $X_t$ is obtained, yielding
\begin{align*}
P(m_t \ge a_t, M_t \le b_t ,X_t \le x_t) &= \int_{-\infty}^{x_t}
\sum_{n=-\infty}^{\infty} [p(z - 2n(b_t - a_t)) - p(z + (2n(b_t-a_t) - 2a_t)] \dd z
\end{align*}
where $p(z)$ corresponds to the density of a standard gaussian density. To include the drift and non-unit volatility, we can use Girsanov's theorem.  Finally, information about the opening price can be incorporated by translating the Brownian motion, i.e, letting $a_t^{*} = a_t -y_{t-1}$, $b_t^{*} = b_t - y_{t-1}$ and $x_t = y_t -y_{t-1}$, which yields \eqref{eq:jointpdf}.  An alternative, more convoluted proof, can also rely on a more probabilistic version of Kelvin's method of images combined with basic Fourier analysis in section 8 of chapter 2 of \cite{PoSto78}.

\bibliographystyle{bka}
\bibliography{stochvol6}
\end{document}